\documentclass[12pt,letterpaper]{article}
\usepackage{amsmath}
\usepackage{amssymb}
\usepackage{cite}
\usepackage{dcolumn} 
\usepackage{setspace}
\usepackage{array}
\usepackage{psfrag}
\usepackage{graphicx}
\usepackage{endfloat}
\usepackage{braket}
\usepackage{rotating}
\usepackage{subfig}
\newcommand{\beq}{\begin{equation}}
\newcommand{\eeq}{\end{equation}}
\newcommand{\beqa}{\begin{eqnarray}}
\newcommand{\eeqa}{\end{eqnarray}}
\newcommand{\bfig}{\begin{figure}\begin{center}}
\newcommand{\efig}{\end{center}\end{figure}}
\newcommand{\btab}{\begin{table}\begin{center}}
\newcommand{\etab}{\end{center}\end{table}}

\newcommand{\rvec}{\ensuremath{\mathbf{r}}}
\newcommand{\Rvec}{\ensuremath{\mathbf{R}}}

\newcommand{\kvec}{\ensuremath{\mathbf{k}}}

\newcommand{\eqn}[1]{\mbox{Eq.\hspace{1pt}(\ref{#1})}}
\newcommand{\eqs}[2]{\mbox{Eq.\hspace{1pt}(\ref{#1}--\ref{#2})}}

\newcommand{\eqtn}[2]{\begin{equation} \label{#1} #2 \end{equation}}

\newcommand{\mfunc}[2]{#1_{#2} \left[ \rho \right] }
\newcommand{\mmfunc}[3]{#1_{#2} \left[ #3 \right] }
\newcommand{\mmmfunc}[4]{#1_{#2}^{#3} \left[ #4 \right] }
\newcommand{\pot}[1]{v_{\rm #1}}

\def\brp{{\mathbf{r}^{\prime}}}
\def\bxp{{\mathbf{x}^{\prime}}}

\def\bk{{\mathbf{k}}}
\def\br{{\mathbf{r}}}
\def\bx{{\mathbf{x}}}

\def\d{{\mathrm{d}}}
\def\rhor{{\rho({\bf r})}}

\def\rhoi{{\rho_I}}
\def\rhoii{{\rho_{II}}}

\def\rhoir{{\rho_I({\bf r})}}
\def\rhoiir{{\rho_{II}({\bf r})}}
\def\rhojr{{\rho_J({\bf r})}}

\def\sumi{{\sum_I^{N_S}}}

\def\im{{\operatorname{Im}}}

\def\etal{{\it et al.}}

\def\se{{Schr\"{o}dinger equation}}

\begin{document}

\begin{center}
\vspace*{1cm}
{\Large\bf Subsystem Density-Functional Theory as an Effective Tool for Modeling Ground and Excited States, Their Dynamics, and Many-Body Interactions}\\[3ex]

{\large Alisa Krishtal, Debalina Sinha, Alessandro Genova, and Michele Pavanello\footnote{m.pavanello@rutgers.edu}}\\
Department of Chemistry, Rutgers University, Newark, NJ 07102, USA\\
\end{center}

\vfill

\begin{tabbing}
Date:   \quad\= \today \\
Status: \> Submitted to J.\ Phys.: Condens.\ Matter as a Topical Review \\
\end{tabbing}
\newpage
\begin{abstract}
Subsystem Density-Functional Theory (DFT) is an emerging technique for calculating the electronic structure of complex molecular and condensed phase systems. In this topical review, we focus on some recent advances in this field related to the 
computation of condensed phase systems, their excited states, and the evaluation of many-body interactions between the subsystems.
As subsystem DFT is in principle an exact theory, any advance in this field can have a dual role. One is the possible applicability of a resulting method in practical 
calculations. The other is the possibility of shedding light on some 
quantum-mechanical phenomenon which is more easily treated by subdividing a supersystem into subsystems. An example of the latter is many-body interactions.
In the discussion, we present some recent work from our research group as well as some new results, casting them in the current state-of-the-art in this review as comprehensively 
as possible.
\end{abstract}
\newpage
\tableofcontents
\newpage
\section{Introduction}
\subsection{The idea behind subsystem DFT}
The founding idea of subsystem Density-Functional Theory (DFT) is that the electron density of a system can be represented by the sum of the electron densities of a collection of subsystems. Namely,
\eqtn{fde_part}{\rho (\br) = \sum_I^{N_S} \rho_I (\br),}
where $N_S$ is the number of subsystems chosen. 
To ensure $N$-representability, each subsystem electron density is constructed to integrate to a predetermined and not necessarily integer number of electrons, $N_I$, such that 
their sum recovers the total number of electrons, e.g.\ $\sum_I N_I=N_
e$. 

Our research group, as well as others, often mention the above equation as a means to simplify the electronic problem by invoking the principle of divide-and-conquer. In fact, in 
the absence of underlying approximations, the above equation is hardly 
useful. One would be inclined to think that \eqn{fde_part} conveys the idea that a system (supersystem, hereafter) can be calculated more efficiently when the problem is fragmented into smaller sub-problems. However, the subsystem electron densities 
on the r.h.s.\ of \eqn{fde_part} are generally unknown, and so is the supersystem density, $\rho$. Thus, from the point of view of quantum mechanics, a density partitioning should not help.

An important realization is that the density partitioning in \eqn{fde_part} is not unique in the absence of additional requirements (constraints). It is the nature of those 
constraints that will make it possible to apply sensible approximations which 
can result in an efficient computation. For example, if the subsystem electron densities can be constructed such that their overlap is minimized, algorithms able to speed up the 
computation by exploiting  locality of the electronic structure 
\cite{bowl2012} can be applied.

There is another approach which is philosophically orthogonal to what is mentioned above -- it prescribes foregoing the quest to computational efficiency, and pursuing another 
goal. 
Namely, finding the most chemically meaningful partition of the 
supersystem into subsystems. Since the pioneering works of Bader \cite{AIM,bade1980}, Hirshfeld \cite{hirs1977}, and Parr {\it et al.} \cite{parr1989,parr1978} this quest has drawn 
the attention of many research groups \cite{nafz2014,nale2000,
elli2010,guse1981}. Such interest stems from the historical fact that chemists gain chemical insight from properties of atoms (or fragments). Thus, a central problem of chemistry 
should be finding an appropriate way of defining atoms or fragments 
in molecules. This has so far been investigated with \eqn{fde_part} as the starting point. We refer the reader to Ref.\ \cite{nafz2014} for a recent comprehensive review on 
density partitioning techniques, especially those aimed at defining 
chemically and physically optimal subsystems. 
Additionally, Refs.\ \cite{libi2014,jaco2014,gome2012} contain outstanding reviews on the subject of subsystem DFT.

In this topical review, we will touch on two competing concepts regarding the merits of subsystem DFT. They can be broadly summarized as the ability to: (1) reduce the computational scaling, and (2) gain a fundamental understanding of the properties 
of the supersystem that otherwise would be hidden in the underlying complexity of the solution of a single electronic problem. For both of these points, we will present some of the most recent work (both published and unpublished) that our group has 
been producing in the past two years. These include: a new implementation of subsystem DFT in the plane wave basis set, an implementation of nuclear (ions) forces for 
Born--Oppenheimer dynamics, an implementation of real-time subsystem TD-DFT, and a 
formulation of a van der Waals theory based on subsystem DFT. Points (1) and (2) are realized in each of the mentioned developments, although at different levels and with different 
outcomes. Each of the topics considered will be cast in the 
framework of the current state-of-the-art.

This review is organized as follows. First, we will describe the two most common computational avenues to achieve a density partitioning, i.e.\ one 
employing kinetic energy functionals and the other one that does not. As our group focuses on the 
former, we will refer the reader to appropriate references for the latter when needed. 
We will also touch briefly on an alternative embedding trend, which partitions the wave 
function rather than the density. 
Second, we will scrutinize the underlying differences between the currently available 
implementations of subsystem DFT, such as basis sets and the choices of 
algorithms for solving the Self-Consistent Field (SCF) equations. After that, we will discuss the extension of subsystem DFT to the time domain by way of the 
Mosquera-Jensen-Wasserman 
theorem \cite{mosq2013,huang2014} (i.e., the extension to subsystem DFT of the 
Runge-Gross theorem\cite{rung1984}). This will provide us with a discussion of the linear-response and the real-time formalisms.

\subsection{Two flavors of subsystem DFT}
\label{flavors}
In this section, we will introduce two of the most popular methods to go about using \eqn{fde_part}. The first method makes use of the so-called nonadditive functionals to describe the interactions between subsystems. The second method, instead, 
requires the availability of the density of the supersystem and solves for those subsystem densities that reproduce the density of the supersystem exctly through \eqn{fde_part}. 
These two methods are very different in spirit, and here we 
will outline their main differences.

It all starts from defining the Lagrangian that needs to be minimized in order to find the electron density of the supersystem,
\eqtn{DFT}{\mmfunc{\mathcal{L}}{DFT}{\rho} = \mmfunc{E}{HK}{\rho} + \int \pot{ext}{(\br)} \rhor \d\br - \mu \left[ \int \rhor \d\br - N_e  \right].}
Herein, $N_e$ is the number of electrons, $\mu$ the chemical potential, and $\mmfunc{E}{HK}{\rho}$ is the Hohenberg--Kohn functional \cite{hohe1964} as reformulated by Levy 
\cite{levy1979} and Lieb \cite{lieb1983}.
The Kohn--Sham DFT (KS-DFT) reformulation of the above Lagrangian \cite{kohn1965} relies on the one-to-one mapping of $v_s$-representable electron densities (we will only consider these type of electron densities in this review unless otherwise 
stated) on to a set of noninteracting electrons, also known as the Kohn--Sham system, whose wavefunction is described by a set of Kohn--Sham orbitals, $\{\phi_i\}$ (curly brackets indicate a set throughout this work). The orbitals are found by 
minimizing the following KS-DFT Lagrangian 
\eqtn{KSDFT}{\mmfunc{\mathcal{L}}{KSDFT}{\{\phi_i\}} = \mmfunc{E}{HK}{\rho} + \int \pot{ext}{(\br)} \rhor \d\br - \sum_{ij}^{N_e} \epsilon_{ij} \left[ \langle \phi_i | \phi_j \rangle - \delta_{ij}  \right].}
Usually, the above Lagrangian features a partitioning of the HK functional into 
\eqtn{KSHK}{\mmfunc{E}{HK}{\rho}=\mmfunc{T}{\rm s}{\rho}+\mmfunc{E}{\rm H}{\rho}+\mmfunc{E}{\rm xc}{\rho},}
where the electronic Coulomb repulsion energy ($E_{\rm H}$), the noninteracting kinetic energy ($T_{\rm s}$) and exchange--correlation energy ($E_{\rm xc}$) functionals are 
introduced. 

By imposing $\mmfunc{\mathcal{L}}{KSDFT}{\{\phi_i\}}$ to be stationary, and only considering the so-called canonical solution (i.e., a diagonal $\epsilon_{ij}$ matrix), the KS equations are recovered \cite{parr1989}. Namely,
\eqtn{KSEQ}{\bigg(\underbrace{-\frac{1}{2}\nabla^2 + v_{\rm ext}(\br) +v_{\rm H}[\rho](\br) +v_{\rm xc}[\rho](\br)}_{\hat f_{KS}}\bigg) \phi_i(\br)=\epsilon_i\phi_i(\br).}

\subsubsection{Use of nonadditive density functionals: The Frozen-Density Embedding method}
\label{sec:nadd}
In the presence of multiple subsystems, each of the subsystem densities needs to integrate to a predetermined number of electrons, in order to ensure $N$-representability.
For this 
purpose, we now consider the subsystem DFT Lagrangian, 
\eqtn{FDE1}{\mmfunc{\mathcal{L}}{S}{\{\rhoi\}} = \mmfunc{E}{HK}{\sumi \rhoir} + \int \pot{ext}{(\br)} \left[ \sumi \rhoir \right] \d\br - \sumi \mu_I\left[  \int \rhoir \d\br - N_I  \right].}
$\mmfunc{\mathcal{L}}{S}{\{\rhoi\}}$ is equivalent to $\mmfunc{\mathcal{L}}{DFT}{\rho}$ in \eqn{DFT} \cite{parr1989} with one redundancy --- the total electron density is artificially written as in \eqn{fde_part} and the constraint of integration to 
the total number of electrons has been split into $N_S$ constraints, one per subsystem. Of course, this partition has no effect on the result of the minimization, and having 
labeled the subsystem chemical potentials as $\mu_I$ is fictitious at this 
stage, as the subsystem chemical potentials must all be equal to the supermolecular chemical potential \cite{fabi2014,grit2013,sand1951}. 

A computationally practical avenue is found by rewriting $\mmfunc{\mathcal{L}}{S}{\{\rhoi\}}$ in a slightly different way. Namely \cite{weso2006,neug2010a},
\eqtn{FDElag}{\mmfunc{\mathcal{L}}{FDE}{\{\rhoi\}} = \sumi \mmfunc{E}{HK}{\rhoi} + \mmmfunc{E}{HK}{nad}{\{\rhoi\}} + \int \pot{ext}{(\br)} \left[ \sumi \rhoir \right] \d\br - \sumi \mu_I\left[  \int \rhoir \d\br - N_I  \right],}
which once again is equivalent to \eqn{FDE1}. The nonadditive HK functional takes the form
\eqtn{NAHK}{\mmmfunc{E}{HK}{nad}{\{\rhoi\}}=\mmmfunc{E}{\rm xc}{nad}{\{\rhoi\}}+\mmmfunc{T}{s}{nad}{\{\rhoi\}}+\mmmfunc{E}{H}{nad}{\{\rhoi\}}}
where $E_{\rm H}$, $T_{\rm s}$, and $E_{\rm xc}$ are not additive and yield the corresponding nonadditive functionals defined as,
\eqtn{nadd}{\mfunc{F}{} = \mmmfunc{F}{}{nad}{\rhoi,\rhoii,\ldots,\rho_{N_S}} + \sumi \mmfunc{F}{}{\rhoi}, ~~ \mathrm{with}~F=T_{\rm s},~E_{\rm xc},~\mathrm{and}~E_{\rm H}.\vspace{-2mm}}

Although the above formulation involving nonadditive functionals has been introduced only relatively recently \cite{sen1986,cort1991,weso1993}, the idea of splitting a supersystem into subsystems and approach the problem from a DFT prospective is 
much older and dates back to the early attempts in applying the ``statistical model'' \cite{fermi1927,gord1972,mass1955,gayd1970} now more commonly referred to as orbital-free DFT \cite{wang2000,huan2010}.

Minimization of the Lagrangian in \eqn{FDElag} with respect to one subsystem density at a time (keeping the others frozen) yields a set of coupled differential equations. Namely,
\eqtn{FDE2}{\left.\frac{\delta\mmfunc{\mathcal{L}}{FDE}{\{\rhoi\}}}{\delta\rhoir}\right|_{\{\delta\rhojr=0\}}
=\left.\frac{\delta\mmfunc{E}{HK}{\rhoi}}{\delta\rhoir}\right|_{\{\delta\rhojr=0\}}+\left.\frac{\delta\mmmfunc{E}{HK}{nad}{\{\rhoi\}}}{\delta\rhoir}\right|_{\{\delta\rhojr=0\}}+\pot{ext}{(\br)}-\mu_I=0.}

In this review, the use of \eqn{FDE2} is considered the definition of the Frozen-Density Embedding (FDE) method. Such semantical choice stresses the fact that the minimization of $\mmfunc{\mathcal{L}}{FDE}{\{\rhoi\}}$ is carried out with {\it 
partial} functional derivatives as opposed to using full functional derivatives as proposed by Lahav and Kl{\"u}ner \cite{laha2007}. In the density embedding literature, sometimes the FDE acronym is applied to a case where only one subsystem is 
included in the variational problem \eqn{FDE2} \cite{weso1993,weso1996b}, and the term ``subsystem DFT'' is reserved for the self-consistent version (i.e.\ all subsystem densities enter the variational procedure). Throughout this review, we will use 
FDE and subsystem DFT interchangeably, as we view FDE [as defined by \eqs{FDElag}{FDE2}] as one possible algorithm to achieve a subsystem DFT treatment of the electronic problem.

The definition in \eqn{FDE2} can be cast in a KS-DFT scheme by transforming the Lagrangian in \eqn{FDElag} to the following auxiliary KS-like Lagrangian
\begin{align}
\label{KSCEDlag}
\nonumber
\mmfunc{\mathcal{L}}{KSCED}{\{\phi_i^I\}} =& \sumi \mmfunc{T}{\rm s}{\rhoi} + \mmmfunc{T}{\rm s}{nad}{\{\rhoi\}} +\\ 
\nonumber
                                     +& \sumi \mmfunc{E}{\rm H}{\rhoi} + \mmmfunc{E}{\rm H}{nad}{\{\rhoi\}} +\\ 
\nonumber
                                     +& \sumi \mmfunc{E}{\rm xc}{\rhoi} + \mmmfunc{E}{\rm xc}{nad}{\{\rhoi\}} +\\ 
\nonumber
                                     +& \int v_{\rm ext}^I(\br) \rhoir \d\br + \int \left[ \sum_{J,~K\neq I} v_{\rm ext}^K(\br) \rhojr \right] \d\br +\int \left[ 
\sum_{J\neq I} v_{\rm ext}^I(\br) \rhojr \right] \d\br\\
-&\left[ \sum_{(ij)_I}^{N_I} \epsilon^{I}_{ij} \left(\langle \phi_i^I | \phi_j^I \rangle - \delta_{ij} \right) \right].
\end{align}
The external potential has been partitioned into subsystem contributions {\it for convenience}, i.e.\ $v_{\rm ext}=\sumi v_{\rm ext}^I$. In going from $\mmfunc{\mathcal{L}}{KSDFT}{\{\phi_i\}}$ to $\mmfunc{\mathcal{L}}{KSCED}{\{\phi_i^I\}}$, the 
supersystem has been mapped onto a {\it collection} of auxiliary KS systems rather than a single one. 
$\mmfunc{\mathcal{L}}{KSCED}{\{\phi_i^I\}}$ is labeled as Kohn--Sham with Constrained Electron Density (KSCED) in line with the previous literature \cite{weso1997}. 

It is important to  point out at this stage that we made no reference as to what defines the subsystems. All we have imposed is that the subsystems must integrate to a preset number of electrons and be $v_s$-representable. Thus, there is no unique 
solution to the variational problem involving \eqn{KSCEDlag}. Disregarding for the moment the issue of non-uniqueness, imposing $\mmfunc{\mathcal{L}}{KSDFT}{\{\phi_i\}}$ to be 
stationary upon variation of a subsystem orbitals, $\{\phi_i^I\}$, while 
keeping the other orbitals frozen, leads to the following one-electron equations which are given in two {\it equivalent} forms,
\begin{align}
\label{KSCED1}
\nonumber
\bigg( -& \frac{1}{2}\nabla^2 + v_{\rm ext}^I(\br) +v_{\rm H}[\rhoi](\br) +v_{\rm xc}[\rhoi](\br) + \\
       +& \underbrace{\left[\sum_{J\neq I}^{N_S} v_{\rm ext}^J(\br) \right] + v_{\rm H}^{I,nad}[\{\rhoi\}](\br)+ v_{\rm xc}^{I,nad}[\{\rhoi\}](\br) + v_{\rm T}^{I,nad}[\{\rhoi\}](\br)}_{v_{\rm emb}(\br)} \bigg)\phi_i^I(\br)=\epsilon_i^I\phi_i^I(\br),
\\
\label{KSCED2}
\bigg( -& \frac{1}{2}\nabla^2 + v_{\rm ext}(\br) +v_{\rm H}[\rho](\br) +v_{\rm xc}[\rho](\br) +v_{\rm T}^{I,nad}[\{\rhoi\}](\br) \bigg)\phi_i^I(\br)=\epsilon_i^I\phi_i^I(\br).
\end{align}
where the nonadditive potentials are defined as $v_{\rm F}^{I,nad}[\{\rhoi\}](\br)=\frac{\delta\mmmfunc{F}{}{nad}{\{\rhoi\}}}{\delta\rhoir}$. We have also defined in \eqn{KSCED1} the embedding potential $\pot{emb}{(\br)}$ as a potential collecting 
all terms linking the various subsystems.

To summarize, in this section we have introduced the fundamental relationships at the base of the FDE method. In FDE, the term of central importance in the interaction between 
subsystems is the nonadditive kinetic energy functional (NAKE) whose 
exact expression is still unknown. Several approximations are available for it with a common starting point being noninteracting KE functional approximants, $\tilde T_s$ \cite{wang2000,wy2013}. There is computational evidence \cite{nafz2014,bern2008,
humb2013} supporting the idea that the non-uniqueness issue is resolved when employing approximate functionals as they effectively minimize the inter-subsystem density overlap.  


%
\subsubsection{Exact embedding: Partition DFT and Potential-Functional Embedding Theory}
\label{pft}
Levy and Perdew \cite{parr1989,lp1985} first, and then Wu and Yang \cite{yang2002,wu2003} provided a mean to compute the KS effective potential when the input quantity is the electron density. This is achieved by first defining the Wu-Yang functional
\eqtn{wy}{\mmfunc{\mathcal{L}}{WY}{v_{\rm s}}=\mmfunc{T}{s}{\tilde\rho}+\int v_{\rm s}(\br) \left( \tilde\rho(\br) - \rho(\br) \right)\d\br,}
which is to be considered a functional of the KS potential only, as the trial density, $\tilde\rho$, is directly obtained from the KS potential.
Minimizing $\mmfunc{\mathcal{L}}{WY}{v_{s}}$ w.r.t.\ the trial density $\tilde\rho$ recovers the KS equations. Maximizing it w.r.t.\ the trial potential $v_{s}$ (which acts 
as 
a continuous set of Lagrange multipliers) yields the unknown KS potential. 
Thus, the solution of the following coupled equations is sought:
\begin{align}
\label{wy1}
\frac{\delta\mmfunc{\mathcal{L}}{WY}{\pot{s}}}{\delta\tilde\rhor}&=0 \to \bigg(-\frac{1}{2}\nabla^2 + \pot{s}(\br) \bigg) \phi_i(\br) = \epsilon_i\phi_i(\br),\\
\label{wy2}
\frac{\delta\mmfunc{\mathcal{L}}{WY}{\pot{s}}}{\delta \pot{s}(\br)}&=0,
\end{align}
with the understanding that \eqn{wy1} involves a minimization, and \eqn{wy2} a maximization. A number of research groups, each with a different flavor of the theory, have followed the idea of reconstructing the embedding potential of FDE from the 
knowledge of the supersystem density \cite{fux2010,good2010,huang2011,jaco2014,cohe2007a}. The first numerical implementation of this technique is due to Roncero \etal\ 
\cite{ronc2008} for regular subsystem DFT and Jacob \etal\ for a density 
partitioning with capping groups \cite{jaco2008}. To aid our explanations, we subdivide the work of others in this framework into two categories, both requiring the supersystem 
density as input.

The first one prescribes a predetermined partition of the density so that \eqn{fde_part} is achieved from the onset of the calculation \cite{ronc2008,jaco2008,fux2010,fux2011,jaco2014}. Then, the unknown becomes the embedding potential that achieves 
such a partitioning. In the case of only two subsystems, this is sometimes cast in terms of a supersystem density, $\rho$, and a so-called ``frozen'' density, $\rhoii$, 
automatically yielding $\rhoi=\rho-\rhoii$ provided that $\rhor-\rhoiir > 0$ \cite{grit2013}. As the KE potential is equal to the negative of the KS effective potential plus a 
constant, the potential reconstruction technique is employed for recovering the effective KS potential for $\rhoi$.  

The second flavor \cite{cohe2007a,nafz2014,tang2012,cohe2009,elli2010,huang2011,huang2011b} poses the problem of finding the most appropriate partitioning of the system. In 
Partition DFT (PDFT),\cite{cohe2007a,nafz2014,tang2012,cohe2009,elli2010} this is done by dividing the total system into predetermined fragments, which can be either atoms, 
functional groups or separate molecules, and minimize the sum of the fragment energies with the restraint that the sum of the fragment densities will match exactly the total molecular 
density. Given the constraint, the total molecular density is an input variable for the PDFT method and the total energy of the supersystem is never calculated. In potential 
functional embedding theory (PFET), the total energy is formulated using the embedding potential $\pot{p}$. This leads to the following 
Lagrangian
\eqtn{PDFT}{\mmfunc{\mathcal{L}}{PFET}{\rho,v_{\rm p}} = \sumi \mmfunc{E}{}{\rhoi} + \int v_{\rm p}{(\br)} \left[ \sumi \rhoir-\rhor \right] \d\br,}
which can be treated in the same way as \eqn{wy}, with $\pot{p}$ in PFET taking over the main role from $\pot{s}$. 
It is quickly verified 
that $\pot{p}$ in \eqn{PDFT} must be equivalent to $\pot{emb}$ of \eqn{KSCED1} 
provided that the embedding potential is imposed to be the same for all subsystems \cite{huang2011}. 
We should mention that PFET and PDFT share \eqn{PDFT} as the main working equation. In both theories the embedding (or partition) potential $\pot{p}$ is unique and shared by all subsystems.

The main drawback of these so-called ``exact'' embedding theories is that solving \eqs{wy1}{wy2} is not computationally efficient.
Despite this, the exact embedding method by Huang {\it et al.} was successfully applied to molecule surface interactions \cite{libi2012,libi2014} in the context of wavefunction-in-DFT. Particularly important is its application to the interaction energy surface of O$_2$ with Al \cite{libi2012} -- a long-standing problem in theoretical surface science.
 We will discuss the computational efficiency and effectiveness of the various flavors of subsystem DFT in section \ref{dac}.

\begin{sidewaystable}
\begin{tabular}{m{5cm}m{5cm}m{3cm}m{3cm}m{2cm}}
Method                   & Unique?                   & Input                                   & Output                         & Central quantity                 \\
\hline
FDE (frozen environment) & Yes                       & $\rho_{\rm frozen}=\rhoii$              & $\rho=\rhoi+\rhoii$            & $\mmmfunc{T}{\rm s}{nad}{\{\rhoi\}}$       \\
FDE (freeze \& thaw)      & No/Yes in theory/practice & $\{N_I\}$                               & $\{\rhoi\}$                    & $\mmmfunc{T}{\rm s}{nad}{\{\rhoi\}}$       \\
PDFT          & Yes                       & $\{v_{\rm ext}^I\}$, $\rho$             & $\{\rhoi\}$, $v_p$, $\{N_I\}$  & $\mmfunc{\mathcal{L}}{PDFT}{\rho,v_p}$ \\ 
PFET             & Yes                       & $\{v_{\rm ext}^I\}$, $\rho$  & $\{\rhoi\}$, $\{N_I\}$, $v_{\rm emb}$      & $\mmfunc{\mathcal{L}}{PDFT}{\rho,v_{\rm emb}}$ \\ 
Pauli blockade           & No                        & $\{N_I\}$                               & $\{\phi_{(i)_I}\}$             & $\sum_{J\neq I, j}|\phi_{(j)_J}\rangle\langle\phi_{(j)_J}|$ \\
\end{tabular}
\caption{\label{tab:gs}Features of the most common subsystem DFT methods.}
\end{sidewaystable}

\subsection{Imposing Orthogonality: A revised Kohn--Sham scheme}
\label{ort}
We have so far considered partitioning the supersystem by \eqn{fde_part}, e.g.\ taking the electron density as the partitioned quantity. An emerging trend is partitioning the supersystem's KS wavefunction. That can be achieved by simply partitioning 
the KS orbitals into subsystems. Namely,
\eqtn{ort1}{\{\phi_i\}\to\bigg\{ \{\phi_{(i)_I}\},\{\phi_{(i)_{II}}\},\ldots, \{\phi_{(i)_{N_s}}\} \bigg\}.}
The above partitioning leads to a trivial implementation of the method if all the orbitals are orthogonal to each other (and not only within a subsystem).
The concept of imposing orthogonality originated in the field of pseudopotentials, originally for the purpose of modelling core electrons and including relativistic effects, but 
also for embedding\cite{bara1988,seij1999,karl2003,swer2008}. An early example of the benefits of imposing orthogonality between the orbitals of different subsystems comes from a 
post SCF calculation presented by Kolos \etal\ 
\cite{kolos1978}. Upon orthogonalization, binding energies calculated with the 
``statistical model'' \cite{gord1972,mass1955,gayd1970} (otherwise characterized by low accuracy) were improved. 
Not imposing orbital 
orthogonality is inconvenient, as the Slater-Condon rules become somewhat 
more complicated \cite{thom2009a,lowd1950}. 
In the KS method, orthogonality is imposed in the Lagrangian, see \eqn{KSDFT}. However, as mentioned before, one might want to extract information peculiar to only one portion of 
the supersystem and in order to achieve that from the onset of the calculation,
it is convenient to split the electronic problem into separate subsystem problems.

In orbital space, such a partitioning can be achieved by the so-called Pauli blockade (PB) method \cite{guto1988} which was recently revived by {\L{}ukasz} \etal\ 
\cite{luka2010b,luka2010}, later by Miller and coworkers \cite{manb2012,barn2013,
good2012}, and Gritsenko \cite{grit2013}, as well as Hoffmann and coworkers \cite{khai2012,tamu2014}. The PB method relies on the effect of a penalty function to be added to the KS Fock operator introduced in \eqn{KSEQ}, which reads
\eqtn{PB}{\hat f_{KS}^{I}=\hat f_{KS} + \sum_{J\neq I}^{N_S} \gamma_J \sum_{j=1}^{N_J} | \psi_{(j)_J}\rangle\langle \psi_{(j)_J} |,}
where $\gamma_J$ are small positive constants.

It is important to remark that from \eqn{ort1} it is clear that the supersystem is mapped onto a {\it single} KS system rather than to a collection of KS systems. This is an 
important point, as it prescribes omitting the NAKE potential in the 
working KS equations. Computational evidence of this can be found in Ref.\ \cite{tamu2014} 
where the penalty in \eqn{PB} was employed correctly (i.e.\ alone) and incorrectly (i.e.\ in conjuction with a NAKE functional).

An overview of the different methods discussed in Sections \ref{sec:nadd}, \ref{pft} and \ref{ort} are summarized in Table \ref{tab:gs}.

\section{Subsystem DFT for the ground state}
There is little doubt that solving the all-electron \se\ is computationally intractable. For example, if a general wavefunction is considered (i.e., not necessarily a Slater 
determinant) the computational cost is $\mathcal{O}(N!)$ with $N$ being the 
size of the system (either the number of electrons or the number of basis functions employed). Avoiding the total solution of the \se\ while maintaining a satisfactory level of 
accuracy of the method is therefore the main goal of electronic structure theory. 
For example, if a Slater determinant alone is employed, the Hartree-Fock method is recovered with an associated formal computational complexity of 
$\mathcal{O}(N^4)$. Unfortunately, this approximation has catastrophic consequences to the accuracy of 
the method and nowadays no scientific publications use Hartree-Fock as the sole electronic structure method. 

DFT was formulated with the goal of finding a method that would provide electronic structures of molecular systems at a cheap computational cost while 
maintaining a satisfactory, or even exact level of accuracy. Since the Hohenberg and Kohn (HK) 
theorems were formulated\cite{hohe1964}, the electron density, $\rho(\br)$, has become the central quantity in electronic structure theory. This marked the birth of 
DFT. The HK theorems stated that once $\rho(\br)$ is known, then in 
principle also the molecular Hamiltonian of the system is available, and with that also the energy functional, $\mmfunc{E}{}{\rho}$. The vice versa is also achievable upon solving 
the appropriate \se. This is a vicious loop -- as in reality neither 
the density nor the energy functional are known, and of course solving the full \se\ is computationally intractable. 

The KS\cite{kohn1965} self-consistent procedure has widened the scope of of DFT by finding yet another direct relationship. This time between $\rho(\br)$ and an 
auxiliary system of noninteracting electrons whose wavefunction is a 
Slater determinant. Again, the Slater determinant is the gate to a computationally cheap method by avoiding tedious and expensive anti-symmetrization procedures. As for the HK 
theorems, KS theory gives no prescription regarding the energy functional 
which is still an unknown. In KS-DFT, the so-called exchange--correlation (XC) functional ($\mmfunc{E}{\rm xc}{\rho}$) needs to be approximated if practical calculations are to be 
carried out. Using local XC potentials yields a method that scales as $\mathcal{O}(N^3)$, and that generally is more accurate than HF. For this reasons, KS-DFT has become the 
standard method for determining electronic structures of molecule and materials.

Because of the $\mathcal{O}(N^3)$ scaling with the system size, KS-DFT is limited in the system sizes it can approach. When modeling a physicochemical process, a careful choice of 
the model system precedes the computations. This is because, systems approachable 
by KS-DFT generally should not exceed 500 atoms, so that the simulations can be completed in a reasonable time. However, it is now understood that if chemical accuracy in the 
predictions is sought, simulations must take into account the complexity 
of the environment surrounding the model system of interest \cite{doum2012,dean2011,bene2013,neug2010}. The continuous struggle to increase the size of the systems in the full 
quantum chemical treatment is facilitated by the development of linear scaling techniques\cite{scus1999,guer1998,gao2004,akam2007} and efficient parallelization 
techniques\cite{gori1997,nees2009}. In Section  \ref{sec:comp} we will show that subsystem DFT enjoys the 
benefit of being linear scaling (provided appropriate approximations are made for the evaluation of the Hartree potential) and fully parallelizable in work and data. Section 
\ref{sec:app} will illustrate some of the applications performed using FDE in recent years, demonstrating its  additional advantage of offering physical insight into the nature of 
the supersystem and the subsystems.

\subsection{Computational complexity and parallelization}
\label{sec:comp}
\subsubsection{A divide-and-conquer approach?}
\label{dac}
For the discussion of the computational speedup associated with subsystem DFT, we will focus on the implementations based on the methods discussed in Section \ref{sec:nadd}, that 
involve the use of nonadditive density functionals. For this flavor of subsystem DFT, the computational gain can be achieved from two sources: (i) reduction of the dimension of 
the size of the problem that needs to be solved (i.e.\ one per subsystem) and (ii) the increased potential to parallelization.

The original paper introducing FDE\cite{weso1993} proposed a simple concept: if the total density can be written as a sum of two densities $\rho(\br)=\rho_1(\br)+\rho_2(\br)$, 
then it is sufficient to minimize the energy of the \textit{total} system w.r.t.\ only one of the densities while keeping the other frozen to obtain the exact 
density and energy of the total system (provided that some mathematical conditions on the choices of the subsystem densities are satisfied). At a first glance, this appears to be an 
uncanny way to reduce the dimensionality of the problem as $N_{sub} < N_{tot}$, except for the following catch $22$: one cannot impose the necessary conditions on the subsystem 
densities without the knowledge of the total density. As a result, one is obliged to perform the exercise iteratively, exchanging the roles of the frozen and variational density, 
also known as the freeze-and-thaw  (FAT) cycles, until convergence is achieved. Of course, even then, one can argue that performing  a calculation of dimension $N_{sub}$ several 
times
is still more computationally advantageous than performing a calculation of dimension $N_{tot}$ once because of the $\mathcal{O}(N^3)$ scaling, but for maximizing the 
computational potential of the method, additional techniques must be applied.

The FAT procedure is designed to improve the initial guess of the frozen density, which needs to be such that at the end of the energy minimization, both subsystem densities (the 
frozen and the variationally minimized) are positive and $v_s$-representable. Numerical tests have shown, however, that in certain cases the 
inclusion of the solvent as a frozen density without relaxation through a FAT procedure, or only a partial relaxation without reaching full convergence, might 
already provide 
satisfactory results \cite{humb2013}. For systems where the full relaxation of all subsystems is necessary, FAT  can be replaced by a much more computationally efficient 
procedure where all subsystems are optimized simultaneously, with a total density update at each SCF iteration \cite{iann2006,shim2001,geno2014}. This is equivalent to 
constructing 
a block-diagonal Fock-matrix, where each subsystem is represented by a block. Both approaches lead to the same minimum if brought to a total point of convergence, where the 
former option might be performed partially and the latter 
option holds a much stronger potential for parallelization, as will be discussed below.

\begin{figure}[tb]
\centering
 \includegraphics[width=\columnwidth]{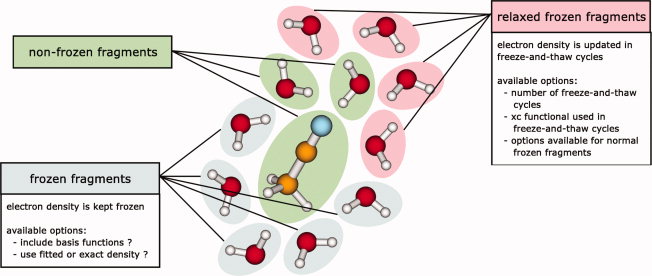}
 \caption{Schematic overview of the fragment-based implementation. The implementation supports nonfrozen fragments, normal frozen fragments and frozen fragments for which the 
density is relaxed in freeze-and-thaw cycles. In addition, a number of options are available for each fragment. Reprinted with permission from ''A Flexible 
Implementation of Frozen-Density Embedding for Use in Multilevel Simulations'',
 C.\ R.\ Jacob, J.\ 
Neugebauer, and L.\ Visscher, J.\ Comput.\ Chem.\ \textbf{29}, 1011-1018 (2008). Copyright (2015), John Wiley and Sons.}
\label{figadf}
\end{figure}

When working with the linear combination of atomic orbitals (LCAO) method for the molecular orbitals, the $\mathcal{O}(N^3)$ scaling is determined by the number of basis 
set functions used in the calculation. A 
straightforward implementation of FDE would involve the use of the supersystem basis set, which would not result in computational speed up but can be useful for 
performing benchmarking calculations \cite{weso1996,jaco2005}.  A more 
computationally efficient option is to use a basis set centered only on the atoms of the 
subsystem in question. Although, depending on the nature of 
the system being studied it might be necessary to include the 
orbitals of other subsystems as well in order to have a basis set flexible enough to model the flow of electron density from one fragment to another \cite{hong2000}. Additionally, 
one can opt for calculating the density of the frozen system using a fitting set instead of using the exact density\cite{jaco2007}  and 
employing efficient numerical integration schemes\cite{neug2005e,shim2001} for large environments. In the subsystem DFT implementation in ADF\cite{jaco2007}, the numerical 
integration grid is centered on the nonfrozen subsystem and therefore, for sufficiently large environments, the integration grid becomes independent of the size of the 
environment 
and, as a result, so does the calculation time of the numerical integration. Another example where subsystem DFT is combined with a numerical integration scheme using hierarchical 
real-space grids is by Shimojo et al.\cite{shim2001}, where, combined with a high level of parallelization, it achieves linear scaling and a parallel efficiency of 0.985 on 128 
processors. The flexible implementation of subsystem DFT in ADF\cite{jaco2007} allows to choose which subsystems will be fully nonfrozen, which will be frozen but relaxed 
using FAT cycles to full or partial convergence, and which will be kept completely frozen (Fig.\ \ref{figadf}). Such a set up is, for 
example, of use 
when studying a molecule in a solvent, where the molecule will be the nonfrozen fragment, the first solvation shell relaxed frozen and the rest of the solvent molecules fully 
frozen.

\begin{figure}[tb]
\centering
 \includegraphics[width=\columnwidth]{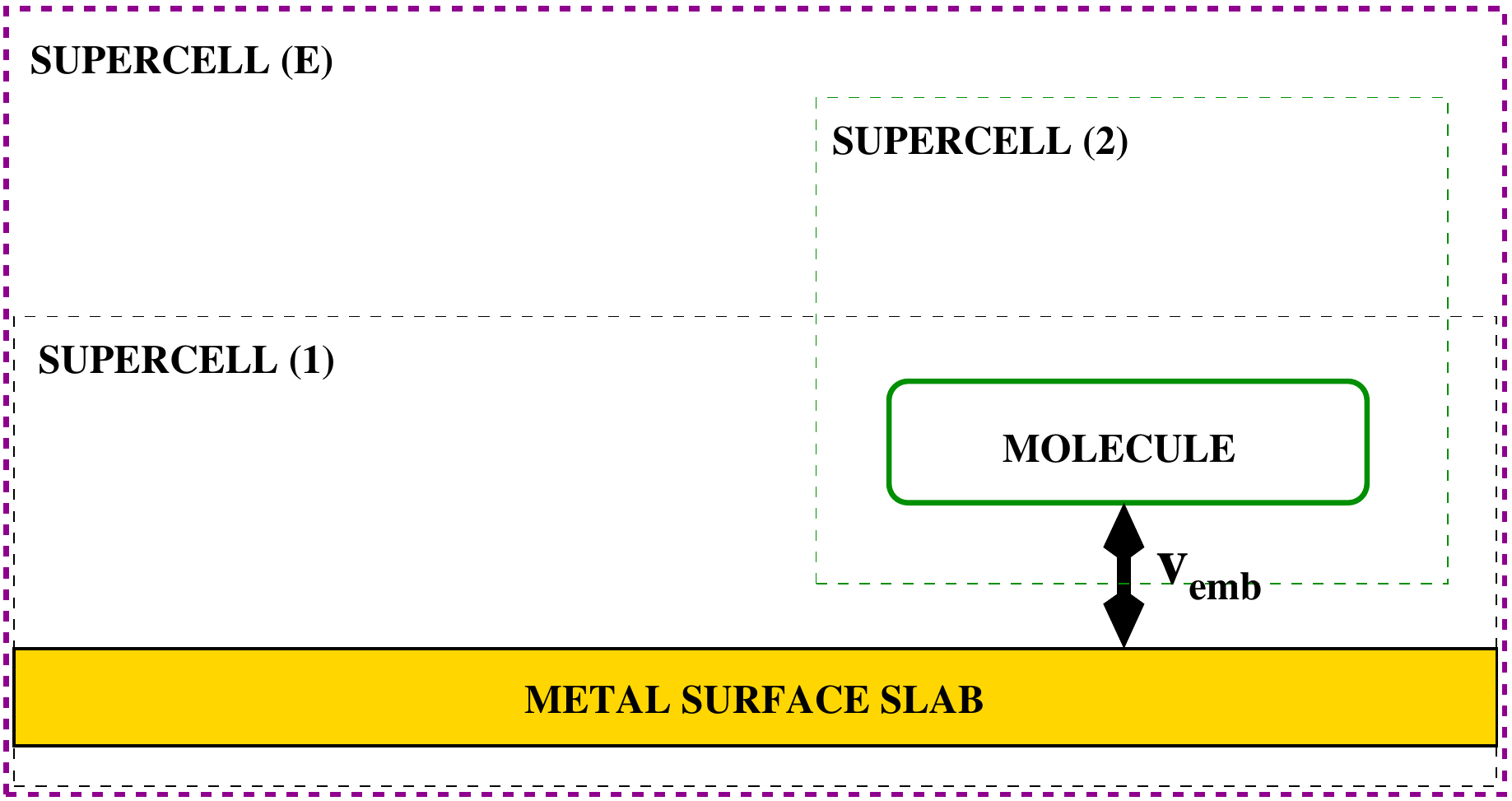}
 \caption{Schematic overview of the interlocking cells concept. The cells of subsystem (1) and (2) are used to evaluate the wave functions of the subsystems, while the 
supercell (E) is used to evaluate the density and the potential.}
\label{fig:interlocking}
\end{figure}

Alternatively, subsystem DFT can be implemented in a code for periodic systems using plane wave (PW) basis sets\cite{geno2014} which are not centered on atoms. Such codes have 
very distinct advantages for applications on periodic systems and condensed phases and have a high potential for parallelization due to the use of a $3$D space 
grid (See for example \cite{mort2005,shim2001,gori1997}). Reducing the dimensionality of the problem in subsystem DFT calculations in PW-based codes is, however, more 
challenging since the basis set size depends on the size of the periodic cell, which, in a straightforward implementation, is determined by the size of the 
supersystem. Reducing the dimensionality in this case is more involved than in codes using localized basis sets, but not impossible. One can implement the use of smaller 
interlocking cells for each subsystem, cut out of a larger supersystem cell. In this case, the KS orbitals of each subsystem can be expanded using the plane waves spanning the 
smaller cells, while the density and the potential are evaluated using the plane waves spanning the supersystem cell. This concept is illustrated in Fig.\ \ref{fig:interlocking}.

\subsubsection{Subsystem DFT: a naturally parallel method}
\label{sec:mpi}

\begin{figure}[tb]
 \centering
 \subfloat[][]{\label{fig:mpi-a}
 \includegraphics[width=.70\columnwidth]{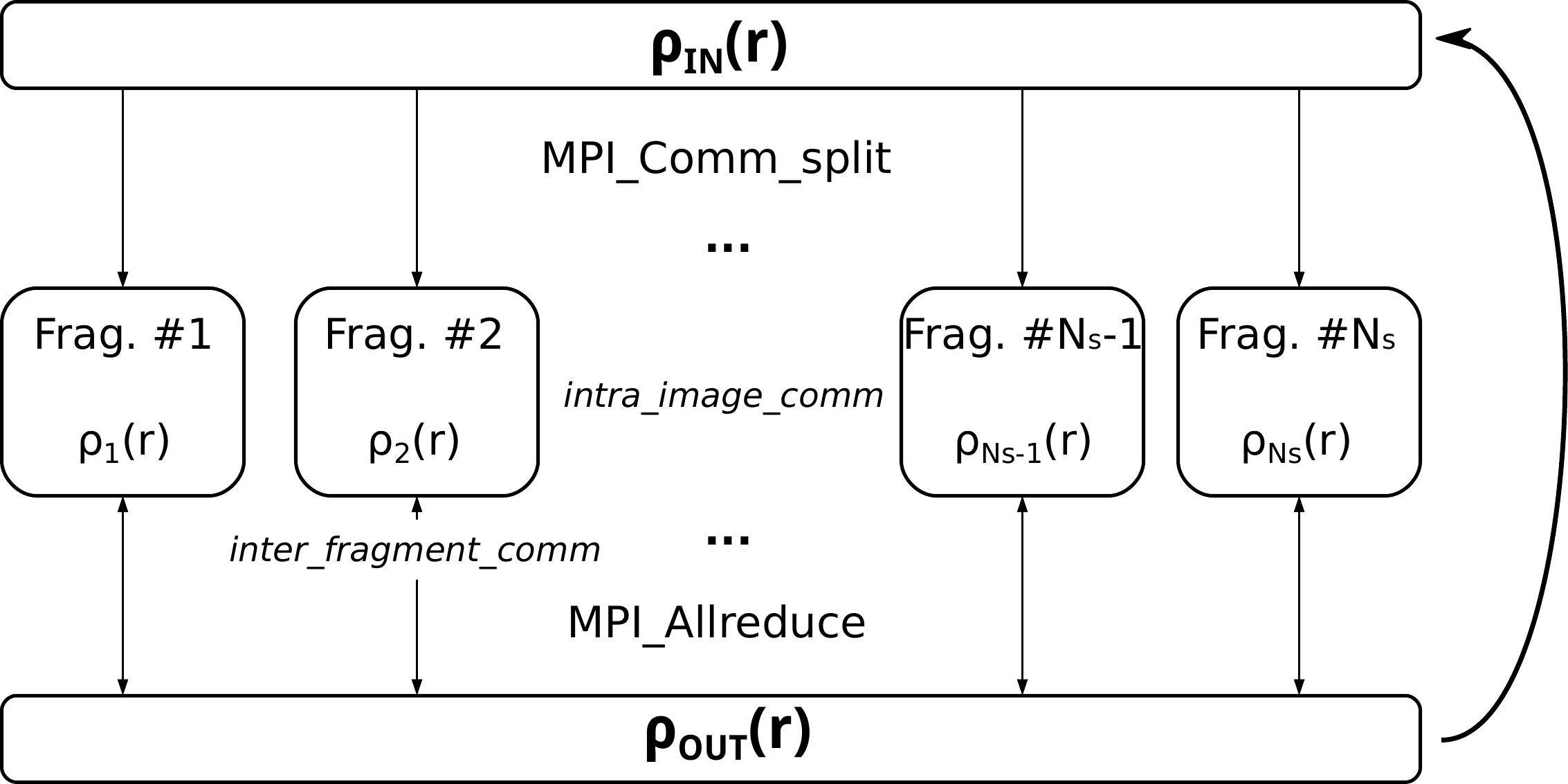}}
 \subfloat[][]{\label{fig:mpi-b}
 \includegraphics[width=.30\columnwidth]{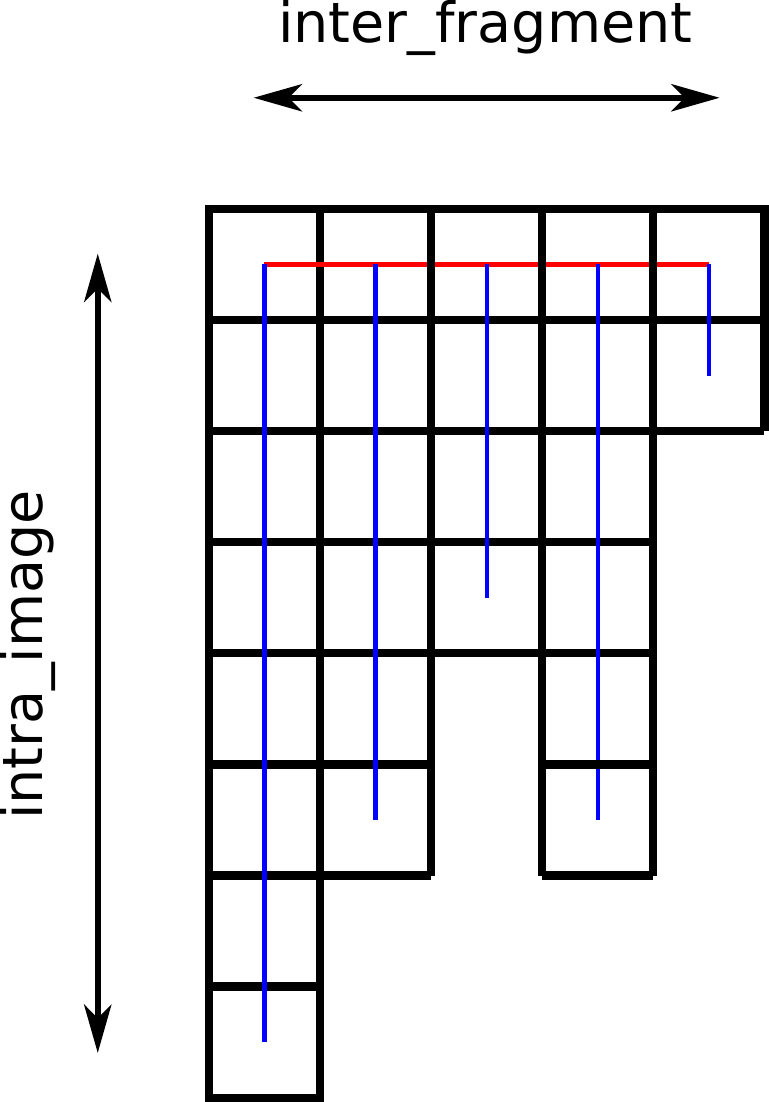}}
 \caption{\textbf{(a):} SCF procedure used in~\cite{geno2014} as alternative to the \emph{freeze-and-thaw} scheme. \textbf{(b):} MPI architecture of a flexible and parallel 
 Subsystem DFT implementation. }
 \label{fig:mpi}
\end{figure}

The efficiency of the subsystem DFT implementation is strongly influenced by the efficiency of the quantum chemical code in which it is implemented. Any advanced computational 
techniques such as linear scaling, numerical integration, nearsightedness and parallelization that are present for the solution of the supersystem KS-DFT problem will be 
transferable to the subsystem DFT problem. Subsystem DFT offers above these optimizations a higher level of parallelization for both work and data. This concept is illustrated in 
Fig.\ \ref{fig:mpi-a} using 
our implementation of subsystem DFT in {\sc Quantum ESPRESSO} \cite{fdeinqe,geno2014}: Each subsystem calculation can be seen as a separate KS-DFT calculation optimizing the KS orbitals of the subsystem. 
In this implementation, the standard option is to update the density between the subsystems after each SCF cycle, though a FAT option also exists. Needless to say, both 
possibilities lead to the same numerical results. At the beginning of each SCF cycle, the total electron density ($\rho_{\mathbf{IN}}(\br)$), composed of the sum of the subsystem 
densities, is communicated to all the subsystems for the evaluation of the potential. At the end of each SCF cycle, the new subsystem densities are reduced to form a new total 
electron density ($\rho_{\mathbf{OUT}}(\br)$). Such scheme is easily implemented using MPI, but one must take care to balance the computation time required by different subsystems 
to perform a single SCF iteration. In the {\sc Quantum ESPRESSO} code this is achieved by optimizing the number of processors assigned to each 
subsystem when the subsystems vary 
in size. During the SCF cycle, the processors assigned to each subsystem are grouped together in an ``intra\_image'' communicator and the code uses the standard parallelization of 
{\sc Quantum ESPRESSO}, where the $3$D  real and reciprocal space grids are distributed between different processors. Between the SCF cycles, the density of each subsystem is 
reduced to an array only allocated on the head process of the communicator, and subsequently reduced and distributed among the subsystems using the ``inter\_fragment'' 
communicator. This concept is illustrated in Fig.\ \ref{fig:mpi-b}.

\subsubsection{Non-additive functional approximants}
We have seen in Section~\ref{sec:nadd} that one way to cast DFT in a subsystem fashion involves the use of nonadditive functionals, i.e., each energy term of the supersystem can 
be 
expressed as the sum of additive and nonadditive contributions. From 
\eqn{nadd} we have
\begin{equation}
\mfunc{F}{} = \sumi \mmfunc{F}{}{\rhoi} + \mmmfunc{F}{}{nad}{\{\rhoi\}} , ~~ \mathrm{with}~F=T_{\rm s},~E_{\rm xc},~\mathrm{and}~E_{\rm H} 
\end{equation}
where the nonadditive term is
\begin{equation}
 \mmmfunc{F}{}{nad}{\{\rhoi\}} = \mmmfunc{F}{}{}{\rho} - \sumi \mmfunc{F}{}{\rhoi} .
\end{equation}

While the above equation is almost trivial for the Hartree and semilocal XC terms, problems arise in the evaluation of the noninteracting KE of the supersystem, when the total 
density is built from the KS orbitals of the subsystems, which are not required to be orthogonal to each other across the fragments.

It is common practice to evaluate the NAKE using pure functionals of the electron density rather than the (unknown) KS orbitals of the supersystem:
\begin{equation}
 \mmmfunc{T}{s}{nad}{\{\rhoi\}} = \mmfunc{\tilde{T}}{{\rm s}}{\rho} - \sumi \mmfunc{\tilde{T}}{{\rm s}}{\rhoi}
\end{equation}
where $\tilde{T}_{\rm s}$ is an approximate functional for the noninteracting KE.

Clearly, the accuracy of the FDE results will be strictly related to the quality of the approximate NAKE functional employed in the calculation. Over the past 
decades 
many KE functionals have been developed. As in the case of the exchange-correlation functionals, 
there are different types of KE functional approximants, spanning from the local TF/LDA functional\cite{thom1927,ferm1928}, to the semi-local GGA 
family\cite{weiz1935,lee1991,lack1994,tran2002,thak1992,lemb1994,kara2006,lari2011}, to fully nonlocal 
functionals\cite{wang2000,huan2010,xia2012b}.

Since the only approximation introduced in the formulation of FDE is the use of approximate kinetic energy functionals for the evaluation of the nonadditive energy 
(and potential), it follows that a systematic way to expand the types 
of systems this method can effectively simulate is to improve the KE functionals. We refer the reader to Ref.\cite{jaco2014} for a comprehensive review on (among others) the NAKE functionals.

\subsection{Applications of subsystem DFT}
\label{sec:app}
\subsubsection{Molecular systems: localized basis sets}
\label{sect:gs}


Since its birth\cite{weso1993}, the FDE method has been implemented in a wealth of Quantum chemical packages. Examples include Amsterdam Density-Functional 
(ADF)\cite{jaco2008b}, TURBOMOLE\cite{lari2010,lari2011}, deMon\cite{
weso1996b,weso1997c}, Dirac\cite{gome2008}, Q-Chem\cite{good2010}, and MolCas \cite{aqui2011a}.
The number of features and the flexibility of the implementations may vary, but all of them share some important similarities, such as the use of localized atom-centered basis sets 
(Slater or Gaussian type). Where available, the FAT scheme\cite{weso1996b} to achieve self-consistency is employed.

The FDE method combined with the currently available semilocal NAKE functionals has been proven very successful in predicting the electronic ground-state properties of a vast array of noncovalently bonded multi-molecular systems\cite{
weso1999,weso1997b,weso1998,weso2001,zbir2004,jaco2006,jaco2008b}. Examples include an accurate prediction of ligand dissociation and proton transfer in metal complexes\cite{hong2000,weso1996c}, the correct interaction energy and electronic 
structure for weak and strong hydrogen-bonding systems\cite{kiew2008,kevo2006}, van der Waals interactions\cite{jaco2005,kevo2014}, and up to a certain extent also 
 reproduced the right electronic structure of Lewis acid-base complexes 
characterized by weak dative bonds\cite{fux2008}.

\begin{figure}[tb]
 \centering
 \includegraphics[width=.5\columnwidth]{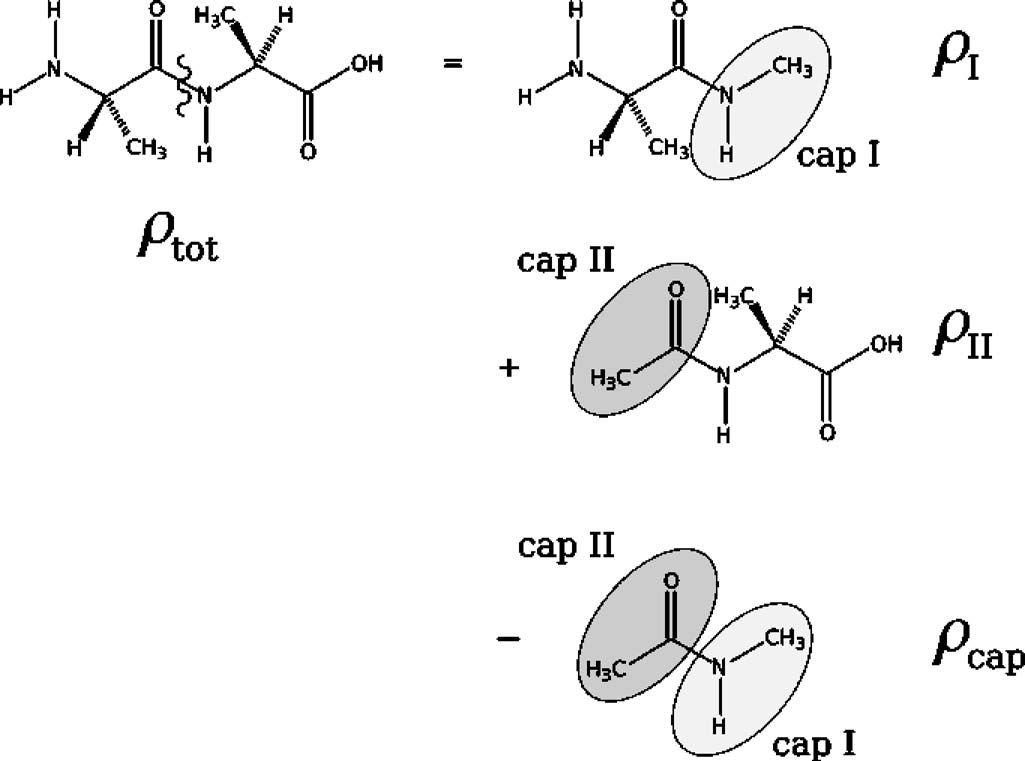}
 \caption{The 3-FDE scheme attempts to allow FDE to simulate fragments connected with covalent bonds, such as polipeptides. The total electron density is $\rho = \rho_I + \rho_{II} 
- \rho_\text{cap}$. Reprinted with permission from ''Density{--}functional theory approach for the quantum chemical treatment of proteins'', C. R. Jacob and L. Visscher, J. 
Chem. Phys. \textbf{128}, 155102 (2008). Copyright 2015, AIP  Publishing 
LLC. }
 \label{fig:jaco2008}
\end{figure}

A powerful flavor of subsystem DFT is 3-FDE. It has been proposed by Jacob and Visscher\cite{jaco2008,kiew2013} to simulate peptide chains and potentially full proteins. Starting 
from the Molecular Fractionation with Conjugate Caps (MFCC) scheme\cite{zhan2003b,gao2004,mei2004}, the chain is cut 
at the peptide bond, and capping groups are added at the extremes to mimic the original environment. The total density of the system is then the sum of the capped amino acids minus 
the density of the capping groups (See Fig.~\ref{fig:jaco2008}). 3-FDE introduced several novelties. The densities are not just those from isolated fragments (as in MFCC), but are calculated using a FDE embedding potential, and possibly converged 
using freeze-and-thaw, it can be applied to subsystems 
connected by covalent bonds, whereas regular FDE would fail due to the approximate NAKE functionals used. 3-FDE has been applied to several peptide pairs and to the ubuquitin protein, yielding 
significantly more accurate results than the MFCC scheme.
More recently, it has been extended to the calculation of excitation energies.\cite{kiew2013,goez2014}

%
\subsubsection{Condensed-phase systems: plane wave basis sets}
\label{planewave}
\label{sec:pw}

Subsystem DFT implementations for periodic systems do not have a history as extensive as the molecular codes counterpart, mostly because the current existing implementations (such 
as CP2K\cite{cp2k,iann2006}, ABINIT\cite{gonz2009,govi1998}, CASTEP\cite{laha2007}) have limitations of different nature, by either limiting the number of fragments to 2, or 
by not 
having the possibility to sample the first Brillouin zone (FBZ).

Efforts in our group to fill this gap are ongoing \cite{geno2014} and involve the implementation of FDE in the plane wave (PW) code Quantum-Espresso (QE) \cite{fdeinqe,qe}.  Using PW 
allows for a very accurate and efficient calculation of the Hartree energies 
and potentials in reciprocal space. Also the evaluation of the forces acting on the nuclei is more straightforward compared to a localized basis set 
implementation (see Sections \ref{sec:bomd} for details and application to molecular dynamics). Moreover, using a PW basis makes the system intrinsically periodic, making it the most natural basis set to expand the Bloch states of such systems, thus 
allowing for a sampling of the FBZ with arbitrary accuracy. Finally, in the subsystem DFT case, we have the option of simulating the fragments using different accuracies in the sampling of the FBZ, potentially leading to additional computational 
time savings.

A disadvantage of a PW implementation is that the use of pseudopotentials in place of the nuclei and core electrons is always needed in order to avoid expanding the one-electron wavefunctions in the fast oscillating nodal region close to 
the nuclei. Specific to subsystem DFT is also the fact that if the same simulation cell is employed for all the fragments, we end up having to solve $N_S$ diagonalization problems 
in the same (large) PW basis set, while localized orbitals 
implementations have the ability of only including the basis functions centered on specific fragments. This means that a speed up in the calculation through a divide and conquer approach does not come as easy for a PW implementation as it does for a 
molecular code (See Section \ref{dac}).

QE offers an ideal platform for the implementation of subsystem DFT, which enables the application of the method to solid state fragments such as surfaces.\cite{fdeinqe,geno2014} 
The FDE implementation is added as a higher level of parallellization to the existing code using the popular Message Passing Interface (MPI) libraries.
To the best of our knowledge, ours is the most comprehensive implementation of subsystem DFT in a PW basis. Specifically, the implementation features simultaneously:
(i) similarly to CP2K and most molecular codes, such as ADF, the ability of simulating an arbitrary number of fragments yielding self consistent electron densities, (ii) a fragment specific sampling of the FBZ (CP2K only samples the $\Gamma$ point, while the other periodic implementations are restricted to only model two subsystems at a time), and (iii) similarly to CP2K, calculation of the energy gradients wrt nuclear displacements. 
To achieve full self 
consistency, our implementation does not use the FAT procedure as a default. Instead, thanks to our MPI machinery and similarly to the CP2K implementation, we are able to overcome 
the active-frozen fragment distinction, simulating all the subsystems at the 
same time. The SCF iterations are synchronized across the fragments and a new embedding potential for each fragment is calculated at each iterative step.
In addition, our implementation allows us to allocate an \emph{arbitrary} number of processors to each subsystem depending on their size (see Section \ref{sec:mpi}).

The formulation of the theory for a periodic implementation of subsystem DFT is very similar to that of the molecular case as it has been explained in Section~\ref{sec:nadd}. However, important distinctions include the fact that we might have 
fractional occupations of the KS orbitals (either to help SCF convergence or to simulate an electron finite temperature), and that the sampling of the FBZ has to be taken into 
account in the equations. As a consequence, the Janak's kinetic energy, 
$T_\text{J}$, of a fragment is found in place of the noninteracting kinetic energy, $T_s$,
\begin{equation}
 T_\text{J} [\rho_I] = \frac{2}{\Omega_\text{BZ}}  \int_\text{BZ} \text{d}\kvec ~ \sum_{j}  n_{j,\kvec}^I \Braket{u_{j,\kvec}^I|-\frac{\left(\nabla+{i}\kvec\right)^2}{2} |u_{j,\kvec}^I},
\end{equation}
where the subsystem Bloch waves are $\phi_i^I(\br)=e^{i\bk\br}u_{i,\bk}^I(\br)$.
The electron density of a fragment with respect to the KS orbitals is given by
\begin{equation}
 \rho_I (\rvec) = \frac{2}{\Omega_\text{BZ}} \int_\text{BZ} \text{d}\kvec ~\sum_{j} n_{j,\kvec}^I\left| u_{j,\kvec}^I(\rvec) \right|^2 .
\end{equation}

\begin{figure}[tb]
 \centering
 \subfloat[][]{\label{fig:pwdd-a}
 \includegraphics[width=.33\columnwidth]{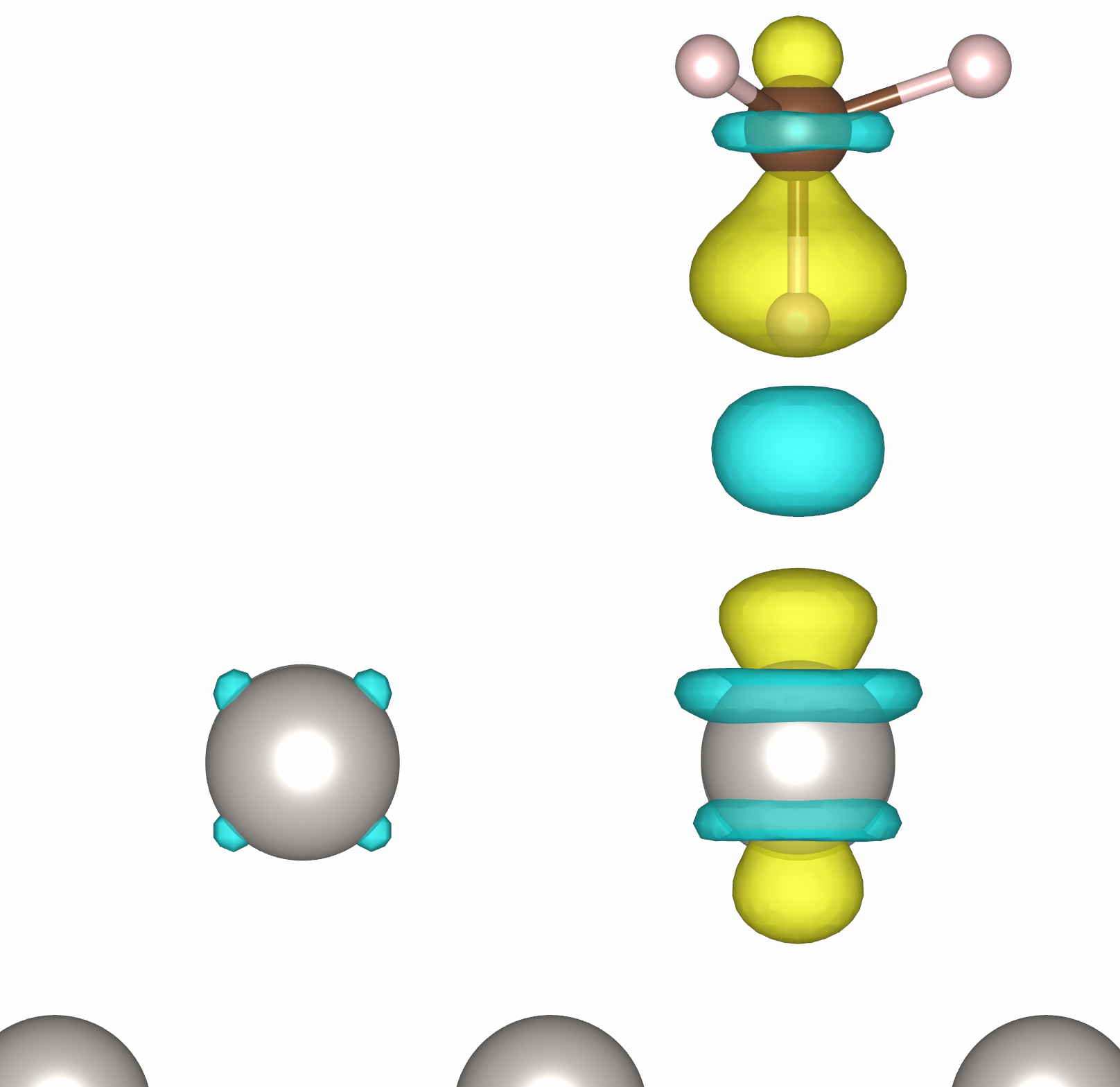}} 
 \subfloat[][]{\label{fig:pwddr-b}
 \includegraphics[width=.33\columnwidth]{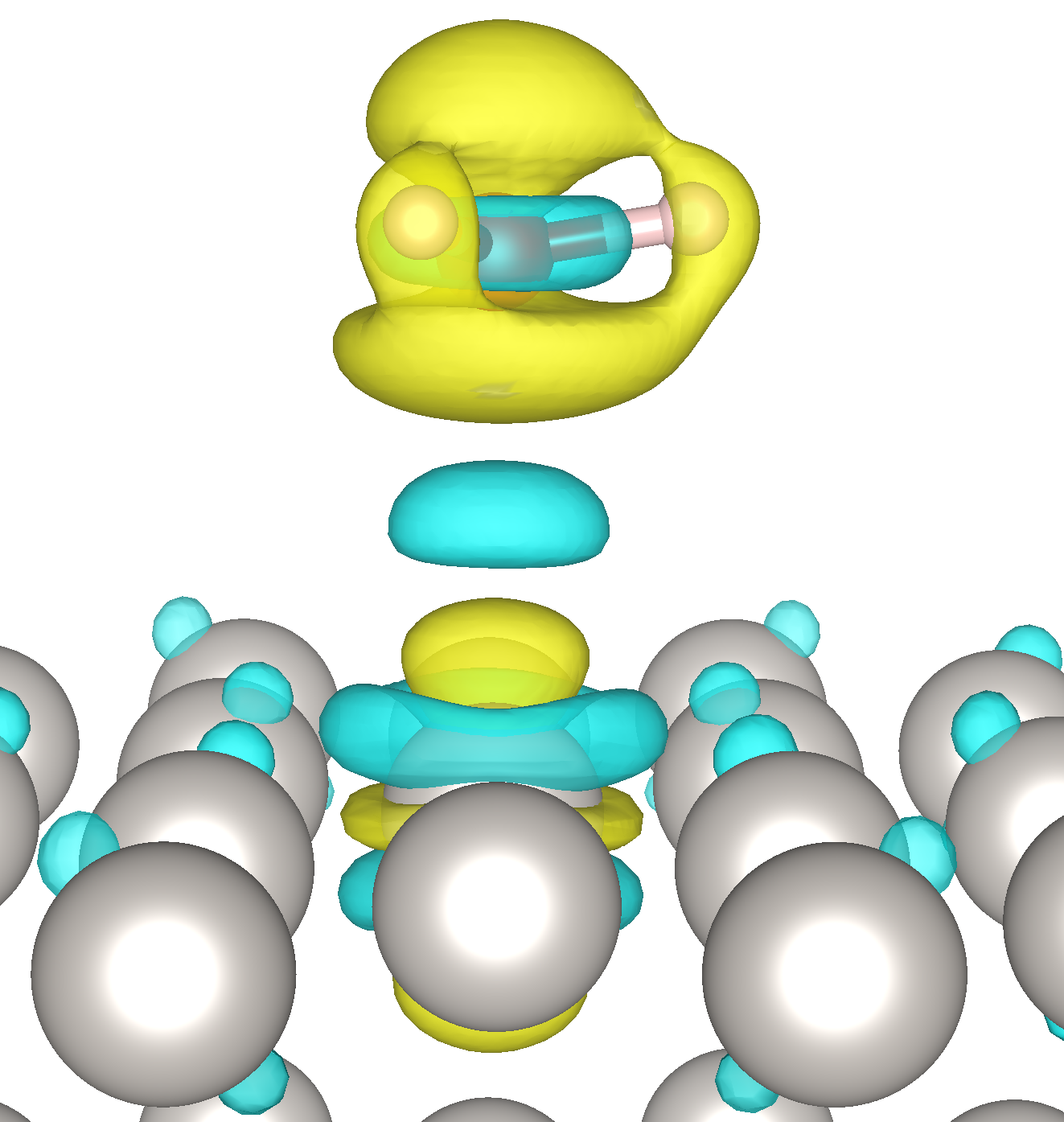}} 
 \subfloat[][]{\label{fig:pwdd-c}
 \includegraphics[width=.33\columnwidth]{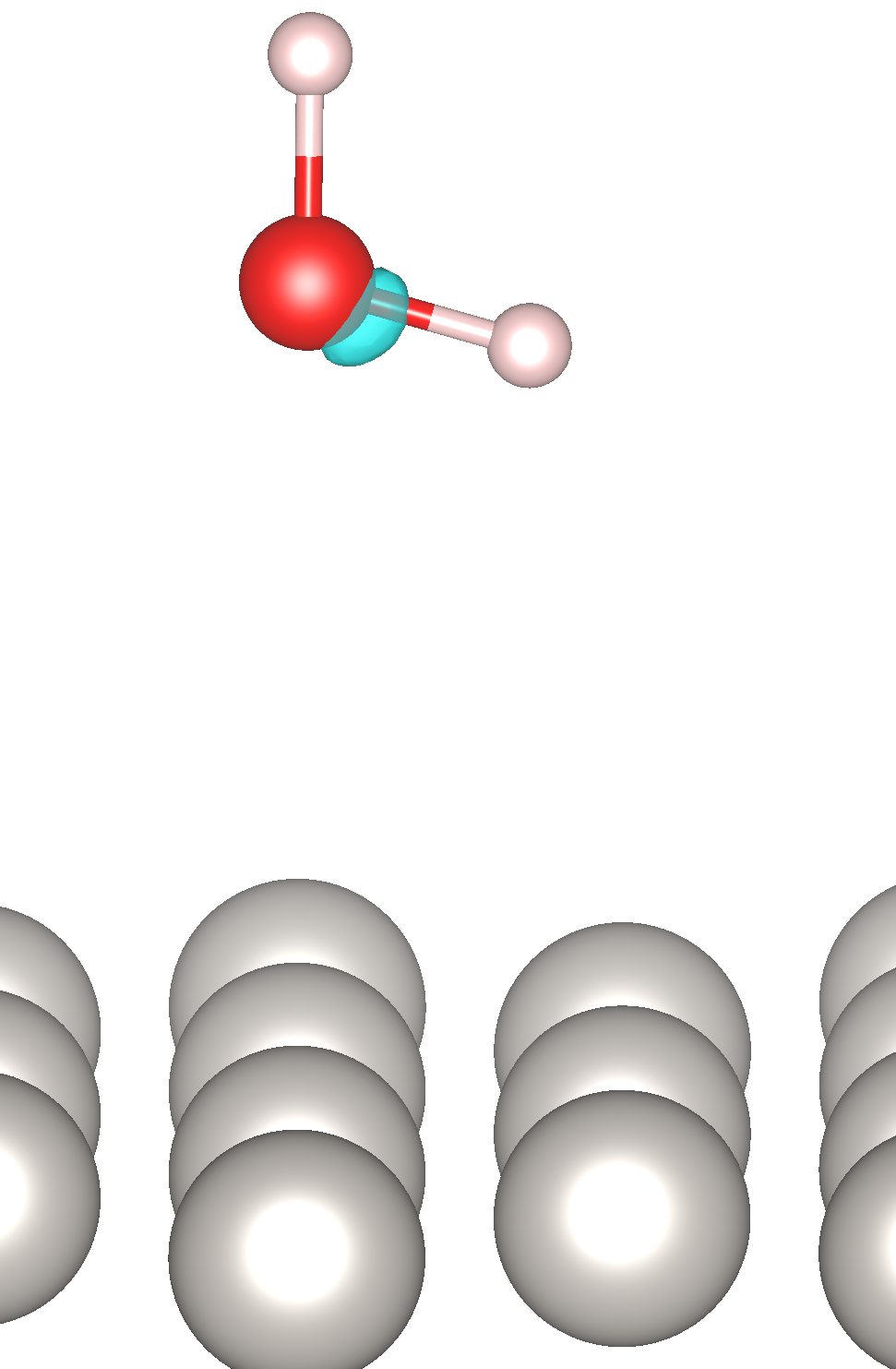}}
 \caption{Isosurface plot of the density difference between a supermolecular KS calculation and a periodic subsystem DFT calculation. \textbf{(a):} Methane on Pt(100), top 
configuration; \textbf{(b):} Water on Pt(111), horizontal configuration; \textbf{(c):} Water on Pt(111), vertical configuration. Reprinted with permission from ''Periodic subsystem density-functional theory'', A.\ Genova, D.\ Ceresoli and M.\ Pavanello, J.\ Chem.\ Phys.\ \textbf{141}, 174101 (2014). Copyright 2015, AIP Publishing LCC..}
 \label{fig:pwdd}
\end{figure}

We proceeded \cite{geno2014} to benchmark our code against systems that had already been studied using localized orbitals codes, observing very similar results: a good agreement with the KS reference for H-bonded systems, and a not completely 
satisfying agreement 
for Lewis acid--base complexes. We then proceeded to study for the first time the interaction of molecules and surfaces using subsystem DFT. Test systems included methane on a 
Pt(100) surface in eight possible configurations, and a liquid water 
bilayer on a Pt(111) surface (a system comprised of 13 fragments). We showed \cite{geno2014} that subsystem DFT carried out with GGA NAKE functionals yields good results when the 
interaction between the molecule and the surface is dominated by 
electrostatics, like for instance the system shown if Figure~\ref{fig:pwdd-c}. On the other hand, due to the known shortcomings of the available GGA NAKE functionals, the quality of the results quickly degrades as the nature of the interaction 
becomes covalent, as it was observed in only two of the possible methane on Pt configurations (Fig.~\ref{fig:pwdd-a}) and by a lesser extent for a parallel water on a Pt(111) 
surface (Fig.~\ref{fig:pwddr-b}).

An important conclusion of our preliminary work is that in employing NAKE functionals with increasing accuracy, the FDE modeling becomes more accurate. 

\subsubsection{Note on self--interaction}
It has been shown \cite{lari2013,solo2014,solo2012} that employing hybrid XC functionals (i.e. functionals including a fraction of the HF exact exchange) for the evaluation of the intra-fragment XC energy, greatly improves the quality of FDE for 
charge-transfer complexes, w.r.t.\ both the interaction energy and the predicted self consistent electron density. This can be rationalized by the fact that hybrid functionals are 
able to reduce the DFT self-interaction error arising from 
the use of approximate exchange functionals, effectively localizing the electron densities of the fragments. When the electron densities of the fragments are less diffused, also the overlap between densities of different fragments is reduced.

The subsystem DFT method has been exploited to constrain spin and excess electron densities on specific fragments \cite{solo2012,pava2011b,pava2013a,ramo2014,ramo2015} even when 
KS-DFT fails to do so due to self-interaction error. This behavior of 
FDE arises from the fact that the FDE theory does not impose orthogonality between the orbitals of different subsystems, thus the typical delocalization of the KS orbitals 
originating from the orbital's hybridization is completely avoided. In 
addition, and probably more importantly, spurious repulsive potential walls arise between the fragments due to the employed semilocal NAKE functionals at the location of the atomic shells of the frozen subsystems \cite{pava2011b,fux2010,ramo2015}. 
It is, therefore, possible to use FDE to approximate the wave function of diabatic states in processes involving charge transfer between a donor and an acceptor fragment \cite{pava2011b,pava2013a,solo2014}, potentially mediated by bridge molecules 
which are also treated in a subsystem DFT fashion \cite{ramo2014}. This framework has recently been applied \cite{ramo2014,ramo2015} with great success to systems of biological interest and realistic size (such as portions of the DNA double helix), 
yielding qualitative and quantitative agreement with experimental data. 

It is important to point out that the diabatic states generated with FDE can only feature charge localization on an entire subsystem. If the diabatic state sought is to feature a charge localization on only a portion of the subsystem, then 
techniques such as constrained DFT \cite{kadu2012} should be coupled with FDE.

\begin{figure}
\begin{center}
\begin{tabular}{c@{\hspace{1cm}}c}
KS-DFT spin-density & FDE spin-density\\
\includegraphics[width=6.0cm]{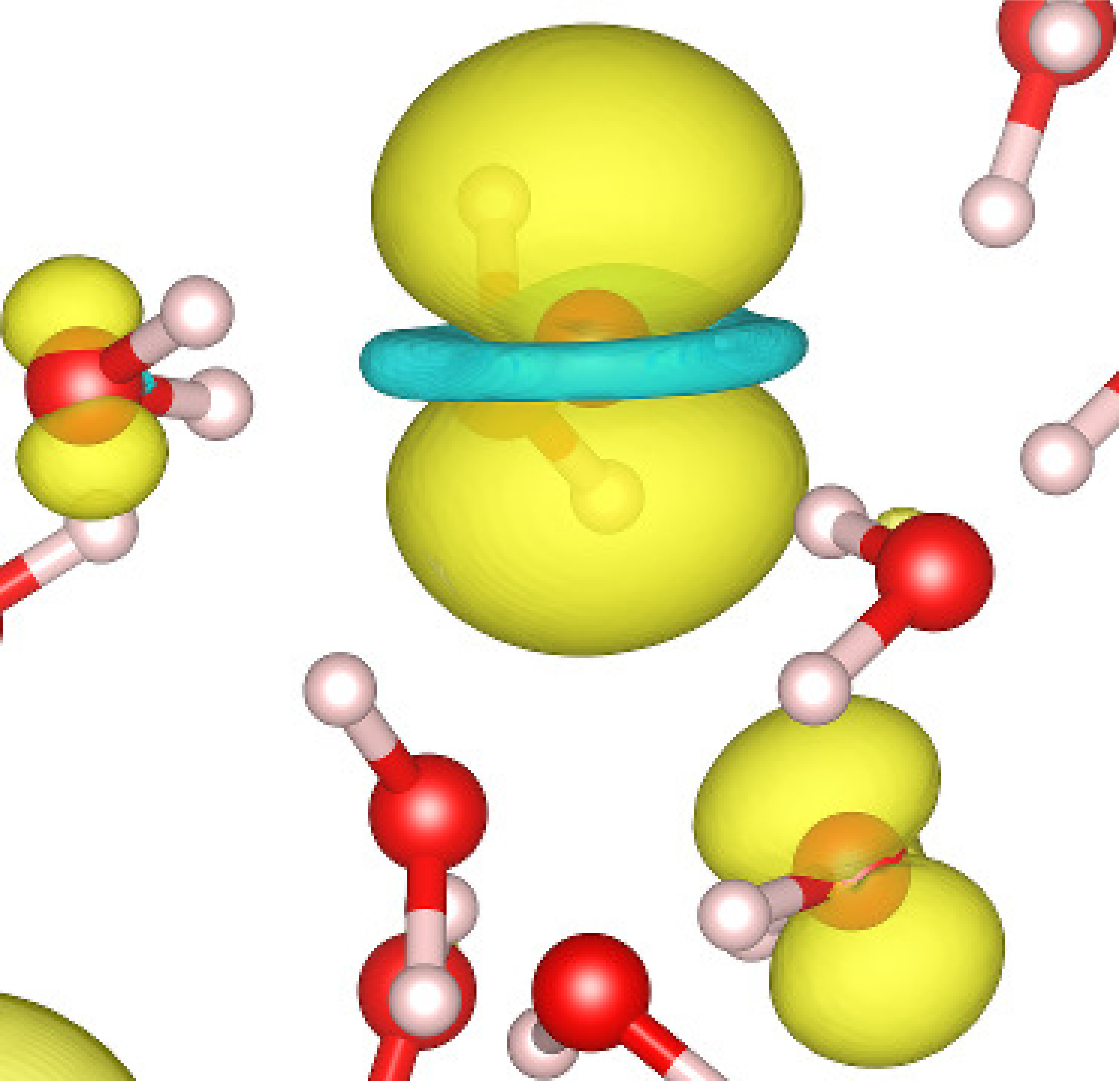} & 
\includegraphics[width=6.0cm]{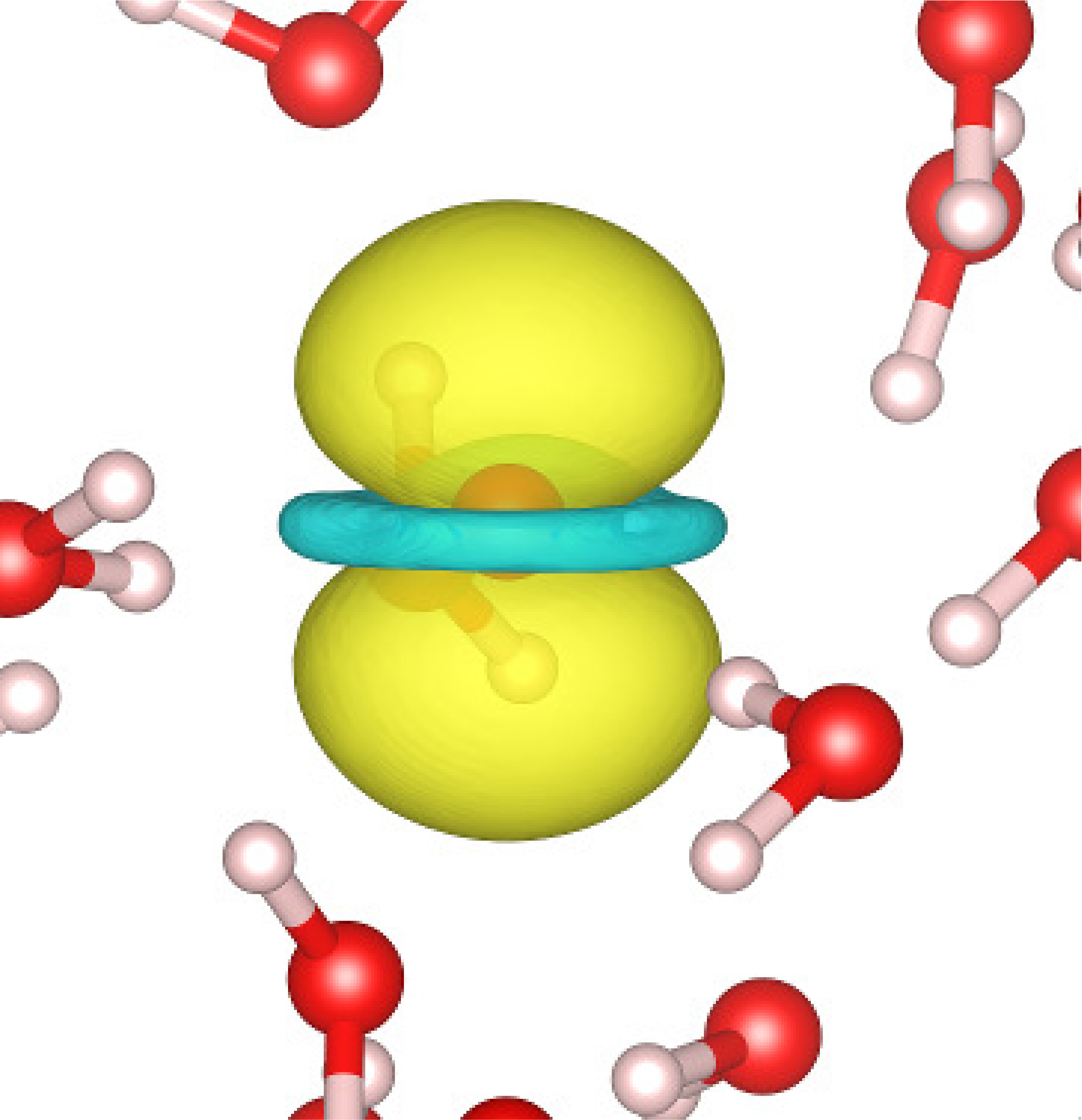}
\end{tabular}
\end{center}
\caption{\label{ohfig}OH$^\bullet$ (H$_{\bf 2}$O)$_{\bf 11}$ carried out with the novel FDE implementation. Spin-density isosurface plot (cutoff of 10$^{-3}$) obtained from FDE and regular KS-DFT are compared. FDE produces a (correctly) localized 
spin-density on the OH$^\bullet$ radical.}
\end{figure}
With our plane wave FDE implementation, we have also ventured into the possibility of localizing charge and spin onto a single subsystem. There are some processes in surface 
science and condensed phase that are not straightforwardly modeled with KS-DFT due to the self-interaction error introduced by GGA XC approximants. Two cases are particularly 
interesting, small molecules interacting with metal surfaces, and solvated 
radical species. When metal surfaces are involved, the simplest reactive 
scattering experiments, such as the ones involving diatomics, have in recent years become theoretical chemistry puzzles [most notably O$_2$+Al(111)] because of the complexity introduced by the self-interaction \cite{carb2010,behl2005,livs2009,
heng2011,libi2012}. With the novel FDE implementation, our group has performed calculations on HO$^\bullet$ embedded in a water solvent. In Figure \ref{ohfig} a snapshot of the 
spin-density extracted from an ab-initio molecular dynamics simulation of 
OH radical embedded in 11 closed-shell water molecules shows that the spin-constrained FDE spin-density is correctly localized on the OH fragment reproducing experimental observations \cite{brau2005} and self-interaction corrected KS-DFT simulations 
\cite{vand2005}. When a semilocal KS-DFT calculation is carried out, the spin-density is too delocalized \cite{vass2004,vand2005} and does not compare favorably with the experiment.

%
%
%
%
%


%

%
\section{Born--Oppenheimer molecular dynamics}
\label{sec:bomd}

A very popular approach when tackling molecular dynamics (MD) is to represent the coupled electron-nuclei time-dependent wave function through the adiabatic approximation
\begin{equation}
 \Phi(\{\rvec_i\} , \{\Rvec_I\};t) = \sum_{n=0}^\infty \Psi_n(\{\rvec_i\};\{\Rvec_I\} \, \chi_n(\{\Rvec_I\};t) \approx \Psi_k(\{\rvec_i\};\{\Rvec_I\} \, \chi_k(\{\Rvec_I\};t)
\end{equation}
where $\Psi_n$ represents the $n$-th solution of the electronic Hamiltonian in the \emph{clamped-nuclei} framework, and $\chi_n$ are time dependent nuclear wavefunctions.

Representing the total wave function as a product of uncoupled nuclear and electronic wave functions allows us to describe the motion of the nuclei by solving a time-dependent 
Schr\"odinger equation involving only the nuclear coordinates, known as the 
Born-Oppenheimer approximation:
\begin{equation}
 \left[ - \frac{1}{2} \sum_I \frac{1}{m_I} \nabla_I + E_k(\{\Rvec_I\}) \right] \chi_k = i \frac{\partial}{\partial t}\chi_k
\end{equation}
where $E_k(\{\Rvec_I\})$ is the potential energy surface (PES) for a given $k$-th electronic state.
Directly solving such equation is not a viable option for most systems, as among other things it would require the knowledge of a $(3N - 6)$ dimensional PES. A popular formulation of ab-initio MD is the Born--Oppenheimer MD (BOMD), where the 
trajectory of the nuclei is propagated classically on a PES calculated on the fly from a ground state electronic structure calculation
\begin{equation}
 m_i\ddot{\Rvec}_I(t) = - \nabla_I \braket{\Psi_0|H_\text{el}|\Psi_0} .
\end{equation}

The forces acting on a nucleus are then calculated using the Hellmann-Feynman theorem, and including Pulay force corrections due to the incompleteness basis set used (in the case 
of an atom centered basis).

If a basis set independent of the position of the nuclei is employed (as plane waves are for instance), there are no Pulay forces. However, in most plane waves calculations 
pseudopotentials (PP) are used to mimic the nuclei and core electrons 
potential. Some pseudopotentials (e.g.\ ultrasoft\cite{vand1990,laas1991,laas1993} and PAW \cite{bloe1994}) contain terms that are determined self-consistently and, at the same time, 
depend on the positions of the atoms. In 
those cases, Pulay corrections to the forces need to be included. Most 
modern PP are split in a local radial part (needed to represent electrostatics) and in a nonlocal term (i.e.\ a term depending on the angular momentum of the single particle 
orbital it is applied to). In a subsystem DFT calculation employing PP, the 
nonlocal part of the PP of atoms belonging to one fragment only apply to the orbitals of that particular fragment, while the local part is shared across the subsystems \cite{iann2006,geno2014}. This ensures that the core electrons are assigned to 
only one auxiliary KS subsystem.

\begin{figure}[tb]
 \centering
 \includegraphics[width=.5\columnwidth]{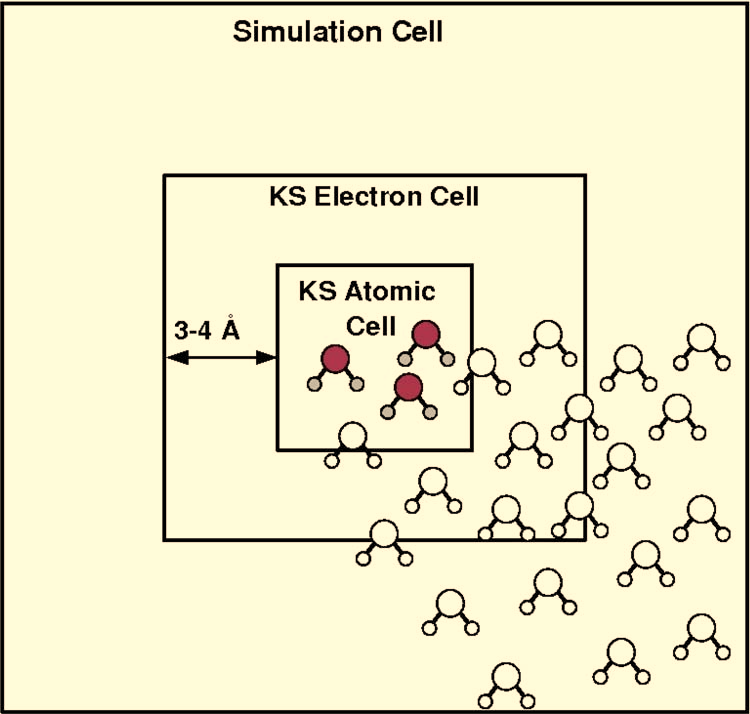}
 \caption{Schematic representation of the hybrid KS-Frozen environment method used by Hodak \emph{et al.} to simulate molecular dynamics of solvated systems. Reprinted with permission from ''Hybrid {\it ab initio} {Kohn}--{Sham} density functional 
theory/frozen-density orbital-free density functional theory simulation method suitable for biological systems'', M.\ Hodak, W\. Lu, and J.\ Bernholc, J. Chem. Phys. \textbf{128}, 014101 (2008)]. Copyright 2015. AIP Publishing LLC.} 
 \label{fig:hoda2008}
\end{figure}

There are not many examples of BOMD based on a subsystem DFT electronic calculations. 
The CP2K implementation of FDE was recently used to model dynamical molecular properties of solvated molecules, with particular focus on the dipole moment \cite{lube2014}.
One study from Hodak and Bernholc\cite{hoda2008} proposed a method combining KS-DFT and a frozen density environment to simulate solvated biological systems. In that study, the 
periodic simulation cell is split in two spatial domains, see Figure \ref{fig:hoda2008}. Atoms belonging to a smaller inner cell (typically a molecule and its first solvation shell) 
are simulated using a full KS method, while the solvent molecules outside this region are modeled by a fixed parametrized electron density: A  
linear combination of Gaussian functions, with one Gaussian function centered on each atom. Along the dynamics simulation atoms can move in and out of the fully quantum mechanical 
cell. Despite the fact that this method is not strictly speaking subsystem DFT, the embedding 
potential of the environment on the QM region is the same as \eqn{KSCED2} (e.g.\ it includes the nonadditive potentials). 

\begin{figure}[tb]
 \centering
 \includegraphics[width=.5\columnwidth]{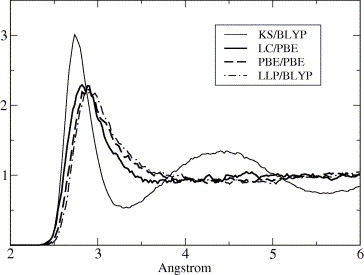}
 \caption{Oxygen-oxygen $g(r)$ distribution for several combinations of kinetic/exchange-correlation functionals as calculated by Iannuzzi \emph{et al.}. Reprinted from ''Density functional embedding for molecular systems'', M.\ Iannuzzi, B.\ 
Kirchner, and J.\ Hutter, Chem.\ Phys.\ Lett.\ \textbf{421}, 16 (2006). Copyright (2015), with permission from Elsevier.}
 \label{fig:iann2006}
\end{figure}

Molecular dynamics of liquid water using FDE have also been studied by Iannuzzi \emph{et al.} \cite{iann2006}. FDE had already been proved capable of predicting the optimal geometry for the water dimer \cite{weso1997b}, but had not been tested on 
liquid water.  Iannuzzi \etal\ \cite{iann2006} carried out BOMD simulations for 64 independent water molecules, using an array of NAKE functionals. It was shown that none of the NAKE functionals available at the time allowed FDE to correctly 
reproduce the experimental radial pair distribution functions, see Figure~\ref{fig:iann2006}. Most notably FDE would fail to reproduce even qualitatively the second solvation shell in the O-O distribution, $g_{O-O}(r)$. A posteriori, this result 
might come as a surprise, because in 2012 Hu \emph{et al.} \cite{hu2012} showed that by using an approximate effective embedding potential fitted from dimer data, a qualitative 
agreement with the experimental radial distribution of liquid water was 
achieved. 

The ability of running molecular dynamics and geometry optimization has also recently been included in our periodic subsystem DFT implementation \cite{geno2014,fdeinqe}. Coupled 
with the ability of the code to sample the FBZ of the fragments, the code 
provides a platform for simulating for example the dynamics of molecules on surface taking advantage of subsystem DFT. 
Current efforts in our group are directed towards applying the code on the dynamics of condensed systems, including shedding light on the elusive liquid water simulations.
\section{Time-dependent subsystem DFT}
\subsection{Extension of the Runge--Gross and van Leeuwen theorems}
\label{rg}

In the ground state, the Hohenberg-Kohn theorem\cite{hohe1964} maps uniquely the wave function to a $v$-representable density, by proving that two potentials differing from each 
other by more 
than 
a constant cannot produce the same density. For the time-dependent case, a similar mapping is done by the Runge-
Gross theorem.\cite{rung1984}
The full quantum chemical description of a time-dependent problem is given by the time-dependent Schr\"{o}dinger equation
\begin{equation}
\label{tdse}
i\frac{\partial}{\partial t}\Psi(t)=\hat{H}(t)\Psi(t),
\end{equation}
where the time-dependent Hamiltonian is partitioned
into a static part and a time-dependent part, which is a direct consequence of the application of an external time-dependent
potential, and could be written as:
\begin{equation}
\hat{H}= {\hat{H}}_0 + {\hat{H}}_1(t)
\label{e2}
\end{equation}
This connects a certain time-dependent external potential $v(\mathbf{r},t)$ to the time-dependent wave function $\Psi(t)$, given an initial state $\Psi_0$. As in the ground 
state, a potential $v'(\mathbf{r},t)$ that differs from $v(\mathbf{r},t)$ by more than an additive time-dependent scalar function $c(t)$ will result in a different wave function. 
The Runge-Gross theorem\cite{rung1984,marq2004} proves that there is an analogical correspondence between the time-dependent potential and the time-dependent density 
$\rho(\mathbf{r},t)$. The situation is however slightly more complicated than in the case of the ground state due to the presence of an initial boundary condition, namely the 
initial state $\Psi_0$. In order to prove the one-to-one mapping between the time-dependent potential and the time-dependent density, one needs to resort to the concept of current 
density 
$j(\mathbf{r},t)$. Without going into the mathematical details, one can show that, when starting from a certain state and two different potentials, the two current densities will 
start diverging from each other at a time infinitesimally larger than $t_0$ and, as a result, also the corresponding time-dependent densities. In other words, the time-dependent 
density is fully determined by the time-dependent potential and the initial state. This also implies that the time-dependent Hamiltonian and wave function are functionals of the 
time-dependent density.

The van Leeuwen theorem\cite{leeu1999,leeu2001,leeu2005} can be seen as the Kohn-Sham theorem equivalent of TDDFT, since it allows us to link a fully interacting system to a 
noninteracting system, resulting in the well known time-dependent Kohn-Sham (TDKS) equations
\begin{equation}
 \label{eq:tdks}
 [-\frac{1}{2}\nabla^2+v_s(\mathbf{r},t)]\phi_i(\mathbf{r},t)=i\frac{\partial}{\partial t}\phi_i(\mathbf{r},t),
\end{equation}
where instead of evolving the full density in time, one evolves the one-particle Kohn-Sham orbitals. We know from the Runge-Gross theorem that a certain time-dependent density 
$\rho(\mathbf{r},t)$, given a certain initial state $\Psi_0$, corresponds to an external potential $v(\mathbf{r},t)$. This system will have a particular electron-electron 
interaction $w(|\mathbf{r}-\mathbf{r}'|)$. van Leeuwen's theorem states that there will exist a different system featuring a different electron-electron interaction  
$w'(|\mathbf{r}-\mathbf{r}'|)$, associated with a different external potential $v'(\mathbf{r},t)$ and a different initial state $\Phi_0$ that will be associated with 
the \textit{same} time 
dependent density. In the special case of 
$w'(\mathbf{r},t)=0$, this is the noninteracting Kohn-Sham system.

Both the Runge-Gross and the van Leeuwen theorems have several practical restriction\cite{ullrich2012} such as the Taylor expandability of the potential and the density, as well 
as the 
$v$-representability of the density. The details of these conditions and their breaking point lie outside the scope of this review. The relevant question here is how the two 
theorems can be viewed in the framework of the 
different flavors of subsystem DFT. 
As discussed in Section \ref{pft}, there are two flavors of subsystem DFT, centered around nonadditive density functionals and around unique embedding potentials. The 
formulation of subsystem DFT influences the way the Runge-Gross and van Leeuwen theorems should be viewed. In Frozen Density Embedding (FDE), the energy is minimized w.r.t.\ the 
density of each subsystem, each of them representing a separate KS system, and therefore also in the TDDFT extension, the Runge-Gross and van Leeuwen theorems have to be discussed 
on a subsystem level. On the other hand, potential functional embedding theory (PFET) and partition density functional theory (PDFT) minimize the energy not w.r.t.\ the 
subsystem density, but with respect to an embedding potential, which is constrained to be a global quantity that is shared by all subsystems. Therefore, the 
Runge-Gross theorem for these methods is formulated with 
respect to the global embedding potential.

At the beginning of a time-dependent FDE calculation,\cite{casi2004,neug2007,krish2014b} each subsystem represents a Kohn-Sham 
system, mapped, on the one hand, to a noninteracting single Slater 
determinant wave function $\Phi^I$ and an effective potential $v_{s}^I(\mathbf{r})$, and on the other hand, a many body wave function $\Psi^I$ and an external potential 
$v^I(\mathbf{r})$. As discussed by Gritsenko\cite{grit2013}, when using approximate kinetic energy functionals for the nonadditive kinetic energy, the external (or effective) 
potential of each subsystem within the FDE method differs from 
the external (or effective) potential of the 
the supersystem
by the error made by $\tilde{T}_s[\rho]$. For each subsystem $I$, the  
external (or effective) potential becomes
\begin{equation}
 v_{ext}^{I}(\mathbf{r})=v_{ext}(\mathbf{r})+\left. \frac{\delta \Delta\tilde{T}_s[\rho]}{\delta \rho(\mathbf{r})} \right|_{\rho=\sum_I\rho_I}-\left. \frac{\delta
\Delta\tilde{T}_s[\rho]}{\delta \rho(\mathbf{r})} \right|_{\rho=\rho_I}
\end{equation}
where $\Delta \tilde{T}_s[\rho]$ refers to the error made by the approximate kinetic energy functional $\tilde{T}_s[\rho]$ with respect to the exact kinetic energy $T_s[\rho]$. 
For the purpose of the Runge-Gross theorem, we will first concentrate on the $\Psi^I$ as the initial state of each subsystem and the time-dependent external potential 
$v_{ext}^{I}(\mathbf{r},t)$. The question that rises is whether one can perform the proof of the Runge-Gross theorem using the \textit{subsystem} external potential to show a 
unique mapping 
between $v_{ext}^{I}(\mathbf{r},t)$ and $\rho^I(\mathbf{r},t)$. The proof of the Runge Gross theorem relies on one major condition, namely the Taylor expansion of the potential 
about the initial time. 
Since the 
subsystem potential differs from the supermolecular potential by the kinetic energy functional error and does not exhibit singularities in time, one can assume this condition holds 
for all subsystems which are part of a supersystem, in which the potential is also Taylor series expandable. For the van Leeuwen theorem, an additional condition is 
necessary: the density must 
be analytic in time at $t_0$.  Usually, it is assumed that when starting from a system that is in the ground state at $t < t_0$, this condition is 
satisfied. However, as shown by Maitra et al.\cite{mait2010}, densities can nonetheless become nonanalytic in time in several special situations. Since all subsystems are in the 
ground 
state at $t < t_0$, we for now assume that this condition is satisfied, but further research regarding the influence of the density partitioning on the analytic time behavior 
of the subsystem densities is needed.

An important result is the unique mapping between the subsystem time-dependent potential $v_{ext}^{I}(\mathbf{r},t)$ and the subsystem time-dependent density 
$\rho^{I}(\mathbf{r},t)$. Since the 
time-dependent potential can be expressed as the sum of the isolated subsystem potential plus and embedding potential [Eq.\ (\ref{KSCED1})], there is also a unique mapping 
between the time-dependent density and the time-dependent embedding potential. In other words, the embedding defines the time evolution of the subsystem in a unique way.

The formulation of the Runge-Gross theorem in the case of the time-dependent PFET\cite{huang2014} and PDFT\cite{mosq2013}
 theories is different, since there the 
time 
evolution is determined by the initial state $\Psi_0$ and the embedding potential $\pot{emb}(\mathbf{r},t)$. In ground state PFET, there is a unique mapping between the total 
electron density and the embedding potential, as opposed to the external potential in DFT. The Runge-Gross theorem has been recast using the time-dependent 
embedding potential first by Mosequra et al.\cite{mosq2013} in the framework of the 
 fragment-based PDFT\cite{mosq2013}
method, and later by Huang et al.\cite{huang2014} in the framework of the 
TD-PFET
method. 
Both proofs show that if a system, defined at $t=0$ by an initial state 
and an embedding 
potential $\pot{emb}(\mathbf{r},t=0)$, evolves into two \textit{different} time-dependent potentials  $\pot{emb}(\mathbf{r},t)$ and  $v'_{\rm emb}(\mathbf{r},t)$ that yield the 
\textit{same} time-dependent 
total electron density, the two embedding potentials can only differ by a time-dependent scalar function. The Runge-Gross theorem can therefore be reformulated using the embedding 
potential.

In practical calculations, the time evolution can be solved either through a linear response mechanism, which will be 
discussed in Section 
\ref{lr-tddft}, or through a direct integration in time of the time 
dependent Kohn-Sham equations [Eq.\ (\ref{eq:tdks})], which will be discussed in Section \ref{rt-tddft}.

\subsection{Linear response approximation}
\label{lr-tddft}

The ground state and excited states of a system which do not evolve with time
are solutions to the time-independent Schr\"{o}dinger equation. In principle, any ground state formalism can be extended easily to obtain
excitation energies by considering slight deviations around the ground state, assuming that the correlation effects dominant in the ground state will also be mostly dominant in 
the excited state.
The search for such a formalism led to the development of the Linear Response (LR) method that describes the response of to an applied 
weak perturbation of a system in its stationary state. Since the perturbation is weak, the response can be truncated at the linear order to a good approximation. The LR method 
can be applied as an extension to any ground state method for energy calculation and when extended in the context of DFT in a time-dependent regime, is known as the 
Linear Response Time-Dependent DFT (LR-TDDFT).
Since its formulation, it has grown to be very popular and have been applied to chemical
problems of varying degrees of complexity \cite{gisb1998,onid2002,marq2004}.

LR-TDDFT allows to obtain excitation energies by performing the following gedanken experiment. Starting from a static system in an initial state, one applies a time-dependent 
external potential of the form $v_{ext}(\mathbf{r},t')=v_0(\mathbf{r})+v_{appl}(\mathbf{r},t')\theta(t'-t)$ and monitors the response of  the system. $v_{0}(\mathbf{r})$ is the 
nuclear potential to which the system is subjected before $t$ and $v_{appl}(\mathbf{r},t')$ is the explicitly time-dependent perturbation turned on at time $t$ using the step 
function $\theta$.
This external potential $v_{ext}(\mathbf{r},t')$ induces a nonstationary state which is described by a linear combination of ground and excited states. Hence it should be possible to obtain the
excitation energies of the system from the structures of the nonstationary wavefunction. 

As discussed in Section \ref{rg}, for each time-dependent density, there is one to one mapping between the fully interacting system evolving under the influence of 
$v_{ext}(\mathbf{r},t)$ and a KS system evolving under the influence of an effective potential $v_{eff}(\mathbf{r},t)$.
The external (or effective) potential at different times $t'$ and positions $\mathbf{r'}$ couples with the density of the system, causing density changes within the system.
For a weak external perturbation, the density change is linear in the potential and the linear response is
given by:
\begin{eqnarray}
\label{eq:deltarho}
\delta \rho(\mathbf{r},t) &=& \int_{-\infty}^\infty dt' \int d\mathbf{r'} \chi(\mathbf{r},\mathbf{r'},t-t') \delta v_{appl}(\mathbf{r'},t') \nonumber \\
&=& \int_{-\infty}^\infty dt' \int d\mathbf{r'} \chi_0(\mathbf{r},\mathbf{r'},t-t') \delta v_{eff}(\mathbf{r'},t')
\label{e5}
\end{eqnarray}
where $\chi(\mathbf{r},\mathbf{r'},t-t')$ is the response function of the fully interacting system and $\chi_0(\mathbf{r},\mathbf{r'},t-t')$ is the
response function of the KS system. 
Now, staring from the von Neumann equation, given by:
\begin{equation}
 i\hbar \frac{\partial}{\partial t} \hat\rho =[\hat H,\hat\rho]
\label{e51}
\end{equation}
in the Heisenberg representation\cite{mcweeny} one obtains 
the response function in the time domain:
\begin{equation}
 \chi(\mathbf{r},\mathbf{r'},t-t')=-i\theta(t-t')\langle\Psi_0|[\hat{\rho}(\mathbf{r},t),\hat{\rho}(\mathbf{r'},t')]|\Psi_0\rangle
\label{e6}
\end{equation}
where $\theta$ is the step function. Applying a  Fourier transformation 
into the frequency and inserting the resolution-of-identity, the Lehmann representation of the frequency dependent density-density response\cite{ullrich2012} is obtained:
\begin{equation}
 \chi(\mathbf{r},\mathbf{r'},\omega)=\sum_{n=1}^\infty \Big{\{} \frac{\braket{\Psi_0|\hat{\rho}(\mathbf{r})|\Psi_n} 
\braket{\Psi_n|\hat{\rho}(\mathbf{r'})|\Psi_0}}{\omega-\Omega_n+i\eta} - \frac{\braket{\Psi_0|\hat{\rho}(\mathbf{r'})|\Psi_n} 
\braket{\Psi_n|\hat{\rho}(\mathbf{r})|\Psi_0}}{\omega+\Omega_n+i\eta}\Big{\}}
\label{e7}
\end{equation}
where $|\Psi_0\rangle$ is the ground state wave-function of the many-body system and $\Omega_n=E_n-E_0$ is the excitation energy.
Eq.\ (\ref{e7}) allows us to extract excitation energies directly from its pole structure: it will have 
poles whenever the frequency
of the probing potential matches with the corresponding excitation (or deexcitation) energy  of a particular excited state $n$. 
However, owing to the time
lag in the application of the external perturbation and the response of the system, a small positive shift $\eta$ appears in the denominator in the complex region of the frequency plane\cite{ullrich2012}.
Thus, the poles of the equation lie in the upper region of the complex frequency plane and its strength depends on the value of the numerator (transition densities).

As $\chi(\mathbf{r},\mathbf{r'},\omega)$ is the density-density response function of a fully interacting system, the poles of Eq.\ (\ref{e7}) have the correct 
structure corresponding to all states. The KS equivalent of Eq.(\ref{e7}) involves the use of single Slater determinants,
which leads to:
\begin{equation}
\chi_0(\mathbf{r},\mathbf{r'},\omega) = \sum_{j,k}^\infty (f_k-f_j) \frac{\phi_k^*(\mathbf{r}) \phi_j(\mathbf{r}) \phi_j^*(\mathbf{r'}) 
\phi_k(\mathbf{r'})}{\omega-(\varepsilon_j-\varepsilon_k)+i\eta}
\label{e9}
\end{equation}
where $f_k$ and $f_j$ are occupation numbers of orbitals $k$ and $j$, respectively. For $\chi_0(\mathbf{r},\mathbf{r'},\omega)$, the non-zero contributions are only when $j$ is an 
occupied orbital
and $k$ is an unoccupied orbital, since all other terms cancel out.
The KS response function only has poles whenever the energy difference $(\varepsilon_j-\varepsilon_k)$ matches the frequency of the applied
external perturbation. Comparing $\chi(\mathbf{r},\mathbf{r'},\omega)$ and $\chi_0(\mathbf{r},\mathbf{r'},\omega)$ reveals an important difference:
The true density-density response function in Eq.\ (\ref{e7}) has multiple solutions of all possible excitation rank while for the KS
density-density response
function, the solutions in Eq.\ (\ref{e7}) are limited to single excitations.
As a result, $\chi_0(\mathbf{r},\mathbf{r'},\omega)$ cannot reproduce the correct pole structure of the fully interacting system. The connection between 
$\chi_0(\mathbf{r},\mathbf{r'},\omega)$ and $\chi(\mathbf{r},\mathbf{r'},\omega)$ is readily extracted from Eq.\ (\ref{eq:deltarho}):
\begin{equation}
\chi^{-1}(\mathbf{r},\mathbf{r'},\omega)= \chi_0^{-1}(\mathbf{r},\mathbf{r'},\omega)-  \frac{1}{|\mathbf{r}-\mathbf{r'}|}-f_{xc}(\mathbf{r},\mathbf{r'},\omega)
\label{e11}
\end{equation}
where $f_{xc}$ is the frequency-dependent exchange-correlation kernel , given by:
\begin{equation}
f_{xc}(\br,t,\br',t')= \frac{\delta v_{xc}[\rho](\br,t)}{\delta \rho(\br',t')} \Big{|}_{[\rho_0]}.
\label{e12}
\end{equation}
The frequency-dependent exchange correlation kernel is the key variable which would generate additional solutions beyond the KS response function and
give correct structures and number of the poles for a true interacting system.
However, since the exact exchange correlation functional is unknown, approximations must be introduced for practical applications.
Among them, there is the adiabatic approximation which involves neglecting the frequency dependence of the exchange correlation kernel, or in other words, 
employing a static exchange correlation potential which
has no dependence on the time-evolution of the entire system at different time $t'$ and position $\mathbf{r'}$. The effect of eliminating the frequency dependence is readily seen 
from Eq.\ (\ref{e9}): any frequency independent shift in the denominator cannot generate additional solutions but can only shift the positions of the poles. Hence, the possibility of accounting of excited states rich in multiple excitations is 
forgone in the adiabatic approximation. Additionally, the 
exchange correlation kernel becomes local in nature with repercussions to the description of states nonlocal in character, such as charge transfer states \cite{mait2005,dreu2005}.

In subsystem DFT, using the idea of LR allows one to obtain the response of a fully interacting system ($\chi$) by calculating the response
of each subsystem ($\chi_I^c$) following simple equation\cite{casi2004}: 
\eqtn{sum}{
\chi=\sum_I \chi_I^{c}
}
where by $\chi_I^c$ we mean a subsystem response function that includes the interaction (coupling) with all other subsystems. The couplings, which are manifested by changes in the densities of subsystems induced by the changes
in the densities of other subsystem, hold valuable physical insight on the nature of the subsystems and the embedding, unaccessible from a supersystem calculation. 

In the framework of subsystem TDDFT, the effective potential acting on each subsystem is given by:
\begin{equation}
 \delta v_{eff}^I(\mathbf{r},t)= \delta v_{appl}(\mathbf{r},t) + \delta v_{ind}^I(\mathbf{r},t)
\label{l11}
\end{equation}
and in the frequency domain for subsystems, Eq.(\ref{eq:deltarho}) transforms to:
\begin{eqnarray}
\delta \rho_I(\mathbf{r},\omega) &=& \int \chi^c_I(\mathbf{r},\mathbf{r'},\omega) \delta v_{appl}(\mathbf{r'},\omega) d\mathbf{r'} \nonumber \\
&=& \int \chi_I^0(\mathbf{r},\mathbf{r'},\omega) \delta v_{eff}^I(\mathbf{r'},\omega) d\mathbf{r'}
\label{l10b}
\end{eqnarray}
where $\chi^c_I(\mathbf{r},\mathbf{r'},\omega)$ is the fully coupled response function of subsystem $I$.
The term $\delta v_{ind}^I(\mathbf{r'},t)$ is a measure of the coupling of the given subsystem
with itself and
with the environment induced by the perturbation. Hence, it is related to the density fluctuations in all subsystems. Namely,
\begin{equation}
 \delta v_{ind}^I(\mathbf{r},t)= \sum_{J}^{N_s} \int K_{IJ}(\mathbf{r},\mathbf{r'},\omega) \delta \rho_J(\mathbf{r'},\omega) d\mathbf{r'}
\label{l12}
\end{equation}
where,
\begin{equation}
 K_{IJ}(\mathbf{r},\mathbf{r'},\omega)= \frac{1}{|\mathbf{r}-\mathbf{r'}| } + 
f_{xc}(\mathbf{r},\mathbf{r'},\omega)+f_T(\mathbf{r},\mathbf{r'},\omega)-f_T^I(\mathbf{r},\mathbf{r'},\omega)\delta_{IJ}
\label{l13}
\end{equation}
where $f_{xc}(\mathbf{r},\mathbf{r'},\omega)$ is the exchange correlation kernel and $f_T(\mathbf{r},\mathbf{r'},\omega)$ and $f_T^I(\mathbf{r},\mathbf{r'},\omega)$ are the 
kinetic energy kernels
of the full system and for each subsystem $I$, respectively. \eqs{l12}{l13} are due to Neugebauer \cite{neug2007}.
The kinetic kernels, when expressed in the time-regime, have the following structure:
\begin{equation}
f_T(\mathbf{r},\mathbf{r'},t-t')= \frac{\delta^2 T_s[\rho]}{\delta \rho(\mathbf{r},t)\delta \rho(\mathbf{r'},t')}
\label{l13a}
\end{equation}
\begin{equation}
 f_T^I(\mathbf{r},\mathbf{r'},t-t')= \frac{\delta^2 T_s[\rho_I]}{\delta \rho_I(\mathbf{r},t)\delta \rho_I(\mathbf{r'},t')}
\label{l13b}
\end{equation}
Using Eq.\ (\ref{l12}) in the expression for density changes, [Eq.\ (\ref{e5})] and transforming to the frequency domain, we arrive at:
\begin{equation}
\delta \rho_I(\mathbf{r},\omega) = \int \chi_I^0(\mathbf{r},\mathbf{r'},\omega) \delta v_{appl}(\mathbf{r'},\omega) d\mathbf{r'}
+  \chi_I^0(\mathbf{r},\mathbf{r'},\omega)\sum_{J} K_{IJ}(\mathbf{r'},\mathbf{r''},\omega) \delta \rho_J(\mathbf{r''},\omega) d\mathbf{r'} d\mathbf{r''}
\label{l14}
\end{equation}
This equation shows how the induced potential couples all density responses of the subsystems with each other.
The coupling in a given subsystem with itself, defined as $\chi_I^u$ is readily obtained as \cite{pava2013b}:
\begin{equation}
(\chi_I^u)^{-1}(\mathbf{r},\mathbf{r'},\omega) = (\chi_I^0)^{-1}(\mathbf{r},\mathbf{r'},\omega) - K_{II}(\mathbf{r},\mathbf{r'},\omega)
\label{l15}
\end{equation}
Inserting Eq.(\ref{l15}) in Eq.(\ref{l14}) we get:
\begin{equation}
\int (\chi_I^u)^{-1}(\mathbf{r},\mathbf{r'},\omega) \delta \rho_I(\mathbf{r'},\omega)d\mathbf{r'} = \delta v_{appl}(\mathbf{r},\omega)
+ \int \sum_{J\ne I} K_{IJ}(\mathbf{r},\mathbf{r'},\omega) \delta \rho_J(\mathbf{r'},\omega) d\mathbf{r'}
\label{l16}
\end{equation}
If the changes in subsystem density are expressed in terms of the coupled subsystem response function and the applied potential, governed by Eq.(\ref{l10b}),
one arrives at a Dyson-type equation given by:
\begin{equation}
\chi_I^c = \chi_I^u + \sum_{J\ne I}^{N_s} \chi_I^u K_{IJ} \chi_J^c 
\label{l17}
\end{equation}
where, if the following expression for $\chi_I^u$ is used:
\begin{equation}
\chi_I^u = \chi_I^0 + \chi_I^0 K_{II} \chi_I^u 
\label{l18}
\end{equation}
we obtain:
\begin{equation}
\chi_I^c = \chi_I^0 + \chi_I^0 \sum_{J}^{N_s}  K_{IJ} \chi_J^c 
\label{l19}
\end{equation}
Eqs.\ (\ref{l17})-(\ref{l19}) connect the total, uncoupled and KS responses for the subsystem. It is important to point out that in the derivation of these equations, no 
further approximations where employed and, in principle, the subsystem LR-TDDFT method is applicable to any system, provided the ground state densities can be correctly described 
using subsystem DFT with the presently available NAKE functionals.

\subsubsection{Applications using subsystem LR-TDDFT}

The original subsystem LR-TDDFT introduced by Casida and Wesolowski\cite{casi2004} offered an approximated but efficient way to include the effect of embedding into the 
the calculation of frequency dependent properties but calculating the \textit{uncoupled} response [Eq.\ (\ref{l18})] of each subsystem, which includes the couplings of the 
excitations within the subsystem but not with its surroundings. When applied to systems where the molecular orbitals are localized on the same molecule, this proved to be a very 
good approximation: for the calculation of excitation energies for guanine--cytosine (G--C) and adenine--thymine (A--T) base pairs in the Watson-Crick arrangements, the uncoupled 
version of subsystem LR-TDDFT was able to reproduce excitation energies differing by less than $0.05$ eV compared to supramolecular calculations.\cite{weso2004}. Additionally, the 
authors were able to extract interaction induced shifts of the monomers compared to the isolated molecules, which ranged between $-0.17$ eV and $0.32$ eV. It is interesting to 
note that the presence of hydrogen bonds in the dimers did not have an effect on the accuracy of the method: the difference between $\chi^u$ and $\chi^c$ is not so much affected 
by the presence of bonds, but by a direct entanglement of the excitations between the subsystems.

The uncoupled version of subsystem LR-TDDFT, from now on labeled FDEu, proved to be very useful for the evaluation of solvatochromic effects, where the presence of a solvent 
results in a shift of an absorption or emission band.\cite{neug2005d} When explicit interactions between the solvent and solute exist, for example in the form of hydrogen bond 
interactions, such effects are known to be problematic to model using implicit models, which only include the dielectric medium effects\cite{mennucci_book}. Using FDEu one can 
apply even a further approximation, by restricting the orbital space to the relatively smaller solute system, while including the solvent only as a 
frozen density for both the ground state and time dependent FDE calculation. In such cases FDEu also offers the advantage of only supplying the excitations of the 
subsystem 
in interest, where in supramolecular calculations the calculation cost is driven further up by the large number of excited states that needs to be included. For 
further reduction of the computational cost, one can generate the density of the solvent as a superposition of spherically symmetric atomic charges, though it has been shown that 
this method can lead to large errors when the molecules within the environment interact strongly by, for example, forming hydrogen-bonded chains.\cite{humb2013} In such cases it 
is recommended to first perform a supermolecular calculation at a lower level of theory. An application on the calculation of excitations of an acetone molecule solvated in water, 
it was found that FDEu was able to produce the correct trend for the energies of the valence excitations with increasing cluster size and the solvatochromic shift 
was found to be in excellent agreement with experiment.\cite{neug2005b} An extensive analysis of the structural effect due to the dynamics in solution and the electronic effect 
due to the frozen densities was performed on a study of aminocoumarin C151 solvated in water and $n$-hexane\cite{neug2005e} showed that the electronic effect is more 
important and a relaxation of the subsystem densities using freeze-and-thaw cycle can be important. In particular, polarization of the density becomes important 
in cases where there is direct hydrogen bonding between the frozen and embedded subsystems, as in case of water. 

While FDEu proved to be successful on many occasions, it also raised the question of when $\chi^u$ and $\chi^c$ start to diverge sufficiently for FDEu to fail.\cite{neug2006b} 
Indeed, in some cases the couplings between the subsystem responses can become dominant. An evident example is a pair of coupled chromophores, where excitation energy transfer 
occurs. If the two chromophores are identical, one can extract information about the excitation rate transfer from the transition-dipole-transition-dipole interaction, which 
expresses itself in the optical absorption spectrum as a splitting of the coupled excitation energy. For this purpose, the subsystem TDDFT version was extended\cite{neug2007} to 
include the calculation of $\chi^c$, between selected intersubsystem excitations. Even when all couplings are taken into account, the dimension of the total problem is still 
smaller 
than in a supermolecular calculation since there are no orbitals defined for the supersystem and hence no inter-subsystem orbital transitions. However, the computational advantage 
of FDE is taken further into account by introducing further approximations. One first solves the FDEu problem, which will contain most of the information on the environmental 
effects due to the embedding potential. One can proceed by only calculating $\chi^c$ for the exciton-like couplings between local transitions.\cite{neug2007,koni2012b,koni2012a} 
For the case of benzaldehyde dimer\cite{neug2007}, for intermolecular distances larger than $5$ \AA, where the excitonic coupling is small, both the FDEu and FDEc methods produced 
similar 
results. For the shorter distance of $4$ \AA, however, a difference of $0.15$ eV was observed. It should be noted that FDEc does not include couplings to excitations with partial 
charge-transfer characters, as the transition densities are expressed in terms of the monomer orbitals. The method can be extended to solvent effects, including selected 
solvent excited states.\cite{neug2010,wy2013} The selection of which excitation should be included in the coupling of the FDEc method, the authors used the sum-over-states (SOS)
expression of the polarizability. The selection was made by ranking the excited states of the solvents in decreasing order of contribution to the SOS polarizability and choosing 
the first $k$ states for reproducing the full polarizability until a threshold of for the desired accuracy was achieved. 
Additional properties for which subsystem LR-TDDFT can be applied include polarizabilities and optical rotation parameters which can be used for the calculation of oscillator and 
rotatory strengths\cite{neug2009}, induced circular dichroism,\cite{neug2005d,neug2006b} vibronic spectroscopy.\cite{neug2004b,neug2005,neug2005d}  and resonance Raman 
spectra.\cite{neug2005c} We refer the readers to Ref.\ \cite{neug2010a} for a detailed overview of specifics of the methods and applications.

A recent development is employing the linear response formalism to perform a WF-in-DFT embedding, as opposed to DFT-in-DFT embedding as discussed until now, in order to 
model local electronic excitations.\cite{khai2010,hofe2012,gome2008,gome2012,koni2014,dada2013} The formal foundations to make this possible were laid in Ref.\ \cite{weso2008}, 
where the expression for the local embedding potential as a functional of charge densities was derived in a general case of embedded interacting Hamiltonian. 
An excellent review on the subject can be found in Ref.\ \cite{gome2012}.

\subsection{Real-time subsystem TD-DFT approach}
\label{rt-tddft}

Real-time TDDFT (rt-TDDFT) involves the direct integration of the TDKS equations [Eq.\ (\ref{eq:tdks})] in time. The main advantages of this approach are that it allows to  
model the response of the system beyond the linear response, obtain optical absorption spectra without the use of virtual Kohn-Sham orbitals and study transport properties 
in real time. It can also be extended to full dynamics of the system by also evolving the nuclei in time and for very large systems it even has computational advantages as it 
scales as $\mathcal{O}(N^3)$. A full discussion of the rt-TDDFT method lies beyond the scope of this review and we will concentrate on the most popular approaches for solving 
Eq.\ (\ref{eq:tdks}) and how they can be extended to various flavors of subsystem rt-TDDFT. The rt-TDKS equations are solved in the following steps:
\begin{enumerate}
 \item A ground state KS-DFT calculation is performed, obtaining the KS orbitals $\phi_j(\mathbf{r})$. These orbitals constitute the initial condition for the time evolution of 
the system. In principle, the initial condition is not required to be a ground state, as long as the initial state of the noninteracting system $\Phi_0$ can be represented as a 
single Slater determinant. For instance, one can construct an initial an excited state by promoting one of the electron to a virtual orbital and performing a constrained DFT 
calculation with the new set of occupied orbitals. 
 \item The time dependent effective potential is set up as
 \begin{equation}
  \pot{s}[\rho](\mathbf{r},t) = 
\pot{appl}{(\mathbf{r},t)}+\pot{ext}{(\mathbf{r},t)}+\int\frac{\rho(\mathbf{r}',t)}{|\mathbf{r}-\mathbf{r}'|}d\mathbf{r}'+\pot{xc}[\rho](\mathbf{r},t)
 \end{equation}
where we adopt the adiabatic approximation for the XC functional, which allows us to use the static XC functional from the ground state 
calculation. The choice of the time dependent applied field depends on the nature of the application and will be discussed below.
\item The integration in time of Eq.\ (\ref{eq:tdks}) is performed, usually using the Crank-Nicolson algorithm\cite{crank1996}, by solving the linear set of equations
\begin{equation}
 \left( 1 + \frac{i}{2} H(t+\frac{\Delta t}{2} )\Delta t \right) \psi_j(\mathbf{r},t+\Delta t)=\left(1-\frac{i}{2}H(t+\frac{\Delta t}{2})\Delta t\right)\psi_j(\mathbf{r},t)
\end{equation}
The $\Delta t$ is needs to be taken very small, around 1-2 attoseconds, to ensure numerical stability. The advantage of the Crank-Nicolson scheme is the preservation of the energy 
and 
the orthonormality of the KS orbitals.
\end{enumerate}

Currently there are two flavors of subsystem real time TDDFT implementations, with exact embedding
theories\cite{mosq2013,huang2014} and with FDE\cite{krish2014b}. The first implementation of fragment-based TD-PDFT was introduced by 
Mosequera et al.\cite{mosq2013}, where a formal proof of the unique  mapping between the total time dependent density and the time dependent partition potential is given. As in 
the ground 
state case, the total system is divided into fragments with the restraint that the sum of the fragment densities at each point of space and time equals the total density
\begin{equation}
 \rho(\mathbf{r},t)=\sum_I\rho_I(\mathbf{r},t).
\end{equation}
One starts by first performing a ground state PDFT calculation in order to obtain the initial fragment KS orbitals. The fragment KS orbitals are propagated in time using the 
following time-dependent fragment equations
\begin{equation}
\label{eq:td-pdft}
 i\frac{\partial}{\partial t}\phi^I(\mathbf{r},t)=[-\frac{1}{2}\nabla^2+\pot{s,I}(\mathbf{r},t)]\phi^I(\mathbf{r},t)
\end{equation}
where the effective fragment potential $\pot{s,I}(\mathbf{r},t)$ contains the partition potential $\pot{p}(\mathbf{r},t)$. The partition potential is determined at each time step 
from 
the total density of the system:
\begin{equation}
 \nabla\cdot[\rho(\mathbf{r},t)\nabla \pot{p}(\mathbf{r},t)]=\frac{\partial^2\rho(\mathbf{r},t)}{\partial 
t^2}+\sum_I(\nabla\cdot\mathbf{Q}_{s,I}[\pot{p}](\mathbf{r},t)-\nabla\cdot[\rho_I[\pot{p}](\mathbf{r},t)\nabla \pot{s,I}[\pot{p}](\mathbf{r},t)])
\end{equation}
where $\mathbf{Q}$ is defined using the density matrix $\Gamma$ and the current density $\mathbf{j}$
\begin{equation}
 \mathbf{Q}_{s,I}=-i \mathbf{tr}\{\Gamma_{s,I}(t)[\hat{\mathbf{j}}(\mathbf{r}),\hat{T}]\}
\end{equation}
The calculation of the time dependent partition potential at each point of space thus requires an independent calculation of the total density. For this reason, the total density 
is propagated 
simultaneously with the densities of the fragments.

In the TD-PFET method developed by Huang et al.\cite{huang2014}, the necessity to calculate the total density at each time 
step is omitted, as they derived a formal equation for the time dependent embedding potential from the action formalism
\begin{equation}
\label{eq:tdpdft_formal}
 \frac{\delta A_{tot}[\pot{emb}]}{\delta \pot{emb}{(\br,t)}}=i\Braket{\Psi_{tot,T}[\pot{emb}] | \frac{\delta \Psi_{tot,T}[\pot{emb}]}{\delta \pot{emb}{(\br,t)}}}
\end{equation}
When the embedding calculations are carried out in such a way that all subsystems are treated within TD-DFT using adiabatic local density approximation (ALDA), 
this formal expression can be approximated as:
\begin{equation}
 \frac{\delta A_{tot}}{\delta \rho_{tot}}=\pot{eff}[\rho_{tot}]+\pot{xc}[\rho_{tot}]-\pot{H}[\rho_{tot}]-\pot{ion,tot}-\pot{td}
\end{equation}
Given that $ \frac{\delta A_{tot}}{\delta \rho_{tot}}$ at $\pot{emb}$ equals zero, this allows to calculate the embedding potential iteratively
\begin{equation}
 \pot{emb}^{(n+1)}(\mathbf{r},t)=\pot{emb}^{(n)}(\mathbf{r},t)-\frac{\delta A_{tot}}{\delta \rho_{tot}}
\end{equation}
where $n$ is the iteration number. To perform a TD-PFET calculation, one starts with a ground state PFET calculation to obtain the fragment densities and Kohn-Sham orbitals, as 
well as $\pot{emb}(\mathbf{r})$, which is used as the initial guess for $\pot{emb}{(\br,t)}$ at $t_0$. The fragment Kohn-Sham orbitals are propagated in time using the 
Crank-Nicolson method with a trial $\pot{emb}{(\br,t)}$ and used to construct the fragment and total densities at $t+\Delta t$. The effective potential and total 
Kohn-Sham orbitals associated with the total electron density are then obtained using the time-dependent version of the Zhao-Morrison-Parr method\cite{zhao1993,zhao1994}, which 
involves the solving of a set of coupled equations self-consistently. From these, one can obtain $\frac{\delta A_{tot}}{\delta \rho_{tot}}$ and a new trial 
$\pot{emb}{(\br,t)}$. This process is repeated iteratively at each time step until conversion of the total electron density. In other words, omitting the necessity to perform a 
supramolecular rt-TDDFT calculation as done in TD-PDFT comes at the cost of two nested iterative calculations for obtaining the time dependent density and the effective 
potential.

An alternative method was introduced by Krishtal et al.\cite{krish2014b}, where the subsystem DFT formulation is extended into a real-time subsystem TDDFT by evolving the 
TDKS equations in time for each subsystem
\begin{equation}
\label{eq:tdksI}
 [-\frac{1}{2}\nabla^2+\pot{s}^I(\mathbf{r},t)]\phi_i^I(\mathbf{r},t)=i\frac{\partial}{\partial t}\phi_i^I(\mathbf{r},t).
\end{equation}
The time dependent effective potential of each subsystem is defined as
\begin{equation}
 \label{eq:vsIrt}
 \pot{s}^I(\mathbf{r},t)=\pot{ext}(\mathbf{r},t)+\frac{\delta J[\rho]}{\delta\rho(\mathbf{r},t)} + \frac{\delta E_{xc}[\rho]}{\delta \rho(\mathbf{r},t)} 
+ 
\frac{\delta\tilde{T}_s[\rho]}{\delta\rho(\mathbf{r},t)}-\frac{\delta\tilde{T}_s[\rho_I]}{\delta\rho_I(\mathbf{r},t)}
\end{equation}
and is equal at $t=t_0$ to the $\pot{s}^I(\mathbf{r})$ in the solution of the ground state subsystem DFT. Only $\pot{ext}(\mathbf{r},t)$ contains an 
explicit time dependence while the other terms only depend on time through the density. If all subsystems are propagated simultaneously and the total density is updated at every 
time step, the full rt-TDKS solution is recovered within the accuracy of the NAKE. If the applied external field is sufficiently small for the 
density response not to diverge from the linear response regime, one can draw a parallel to the response functions from linear response TDDFT, discussed in more detail in Sec. 
\ref{lr-tddft}. In that case, the solution of each subsystem includes the full 
coupled response of the subsystem $\chi_I^c$\cite{pava2013b}, as will be discussed below in Section \ref{lr-tddft}. If Eq.\ (\ref{eq:tdksI}) is only propagated for one of the 
subsystems, while 
keeping the other subsystems frozen, the solution will only include the uncoupled subsystem response $\chi^u_I$. The subsystem rt-TDDFT method was implemented into the {\sc 
Quantum Espresso} package using plane wave basis sets, as described in Section \ref{planewave}. The solution of the equations is performed in a similar fashion to the rt-TDDFT, 
by applying the 3 steps described at the beginning of this Section to each subsystem. For the case of the full coupled calculation, the total 
time dependent density is reconstructed from the densities of the subsystems at the end of step 3.

The three subsystem rt-TDDFT methods are summarized in Table \ref{tab:rttddft} w.r.t. main quantities, the required input at the initial time $t_0$, the required input at each 
time step and the output quantities.

\begin{sidewaystable}
\begin{tabular}{m{2cm}m{5cm}m{5cm}m{8cm}m{2cm}}
Method                   & Input at $t_0$            & Input at each $t$                       & Output at each $t+\Delta t$             & Central quantity   \\
\hline
FDE                      & $\{\phi^I(\mathbf{r})\}$, $v_s^I(\mathbf{r})$  & $\{\phi^I(\mathbf{r},t)\}$,  $v_s^I(\mathbf{r},t)$                & $\{\phi^I(\mathbf{r},t+\Delta 
t)\}$   &$\rho^I(\mathbf{r},t)$      \\
PDFT                     & $\{\phi^I(\mathbf{r})\}$, $v_p(\mathbf{r})$, $\rho(\mathbf{r})$      & $\{\phi^I(\mathbf{r},t)\}$, $\rho(\mathbf{r},t)$      
       & $\{\phi^I(\mathbf{r},t+\Delta t)\}$, $\pot{emb}{(\mathbf{r}, t)}$& $\pot{emb}{(\mathbf{r},t)}$ \\ 

PFET                     &  $\{\phi^I(\mathbf{r})\}$, $\rho_I(\mathbf{r})$, $v_{emb}(\mathbf{r})$  & $v_{\rm emb}^{app}(\mathbf{r},t)$, $\rho_I(\mathbf{r},t)$, 
$\{\phi^I(\mathbf{r},t)\}$ &$v_{\rm emb}^{exact}(\mathbf{r},t)$, $\rho_I(\mathbf{r},t+\Delta t)$, $\{\phi^I(\mathbf{r},t+\Delta t)\}$    & $\pot{emb}{(\mathbf{r},t)}$ \\ 
\end{tabular}
\caption{\label{tab:rttddft}Summary of the subsystem rt-TDDFT methods.}
\end{sidewaystable}

\subsubsection{Electronic spectra}

Electronic spectra can be obtained with rt-TDDFT by applying a short laser pulse of strength $\epsilon$ to the Kohn-Sham orbitals of the system, propagating them in time and 
Fourier 
transforming the time-dependent dipole moment $\mu(t)$ into the frequency domain. The pulse can be either applied by shifting the Kohn-Sham states\cite{yaba1996}
 \begin{equation}
\label{eq:pulse}
 \phi_i(\mathbf{r},t=0^{+})=e^{i\epsilon\mathbf{r}}\phi_i(\mathbf{r},t=0^{-})
\end{equation}
where $\epsilon$ is the field strength or by adding an explicit time dependent potential in the form of a very narrow gaussian in time that integrates to $\epsilon$. The second 
option is not feasible in periodic calculations but both 
alternatives produce identical results for molecular systems. If $\epsilon$ is sufficiently small, the density response will be confined to the linear response regime and the 
results will be directly comparable to those obtained the using the linear response TDDFT formalism, discussed in Section \ref{lr-tddft}. By applying a stronger field, one can 
straightforwardly study effects beyond the linear response, which is one of the main advantages of rt-TDDFT. This, however, requires extra care in the numerical time 
integration of Eqs.\ (\ref{eq:tdksI}), such as the use of very small time steps ($<1$ as) and a predictor-corrector scheme for the Crank-Nicholson algorithm. All applications 
reported here were performed using a small electric field.

The oscillator strength at each 
frequency $\omega$ is given by
\begin{equation}
 \label{eq:oscillator}
 S_k(\omega)=-\frac{2\omega}{e_k\pi}\int\sin(\omega t)e^{-\gamma t}[\mu_k(t)-\mu_k(t_0)]dt
 \end{equation}
where $\gamma$ is a small damping factor associated with the $\eta$ factor in Eq.\ (\ref{e7}). 
The resolution $\Delta\omega$ of the spectrum depends on the simulation time and usually a simulation of at least 10 fs is necessary.
At the present, no results on electronic spectra have been reported using the 
TD-PFET and TD-PDFT methods, so we will concentrate on results obtained using subsystem rt-TDDFT method\cite{krish2014b}.

\begin{figure}
\centering
 \includegraphics[scale=0.6]{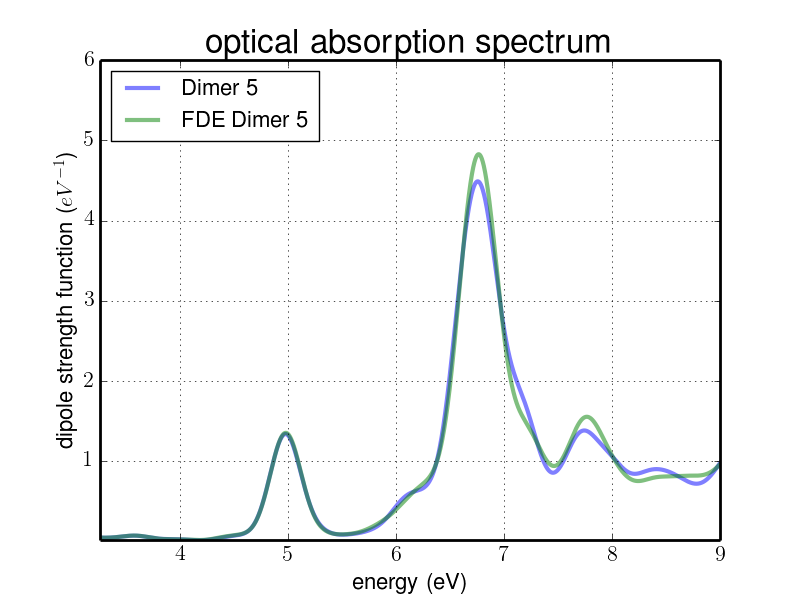}
\caption[]{The electronic spectrum of the stacked benzene-fulvene dimer at the separation distance of $5$ \AA, obtained using the rt-TDDFT and FDE-rt-TDDFT methods.}
 \label{fig:dimer_fde}
\end{figure}

An example of what kind of information that can be obtained from a subsystem rt-TDDFT method is a dimer of stacked benzene and fulvene molecules, separated by a distance of $5$ 
\AA. The optical spectrum is obtained by applying a laser pulse of $0.01$ Ryd \AA\ and evolving the Kohn-Sham orbitals of each of the subsystems in time for $8000$ steps with a 
time 
step of 2 as.  The dimer was placed in 
a supercell of $31.0 \times 32.5 \times 37.8$ a.u.$^{3}$ to avoid periodic interactions. The calculation was performed with the PBE functional\cite{PBEc},  the LC94\cite{LC94} 
functional for NAKE and ultrasoft 
pseudopotentials from the 
GBRV\cite{garri2014} library with a kinetic energy cutoff of $55.0$ Ry and density cutoff of $660.0$ Ry. Fig.\ \ref{fig:dimer_fde} compares the results obtained using rt-TDDFT and 
subsystem rt-TDDFT. The spectrum of the dimer in the subsystem rt-TDDFT  method is obtained by simply adding the spectra of the two subsystems, since the time dependent dipole 
vector is additive across the subsystems. Since benzene and fulvene do not absorb in the visual region, the spectrum is shown for the $3.26$-$9$ eV range. Higher energy 
frequencies 
are not considered here. FDE succeeds in reproducing the TDDFT results with slight differences resulting from the approximate kinetic energy 
functionals used in the embedding potential. All excitations under $8$ eV are reproduced within the $0.01$ eV accuracy, with only slight deviations in the oscillation strength. At 
higher energies region, the deviations are slightly larger: this could also be an attribute of the numerical accuracy of the rt-TDDFT implementation, since higher frequencies are 
reproduced at 
shorter simulation times and must be commensurate to the time-step.

\begin{figure}
\centering
 \includegraphics[scale=0.6]{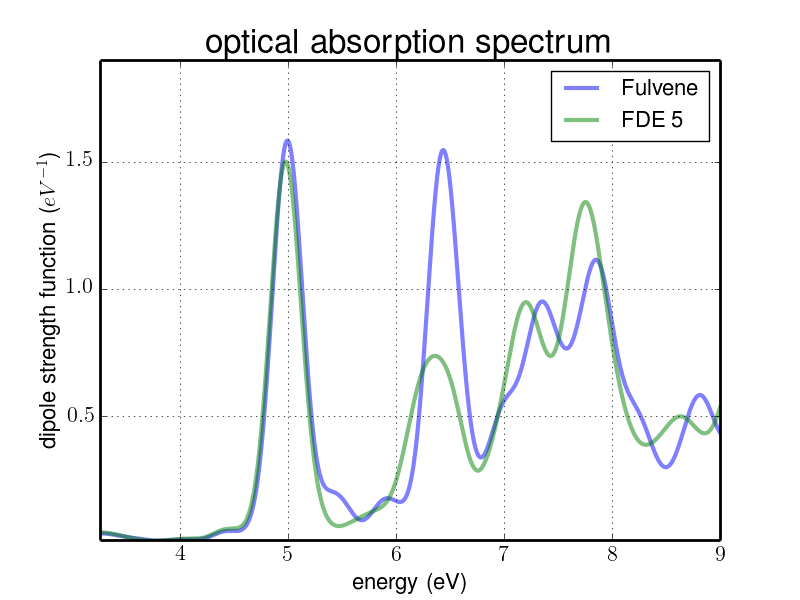}
\caption[]{The electronic spectrum of the isolated fulvene molecule obtained using rt-TDDFT and the fulvene molecule in the benzene-fulvene dimer at $5$ \AA\ separation, obtained 
using subsystem rt-TDDFT.}
 \label{fig:benzene_iso_fde}
\end{figure}

\begin{figure}
\centering
 \includegraphics[scale=0.6]{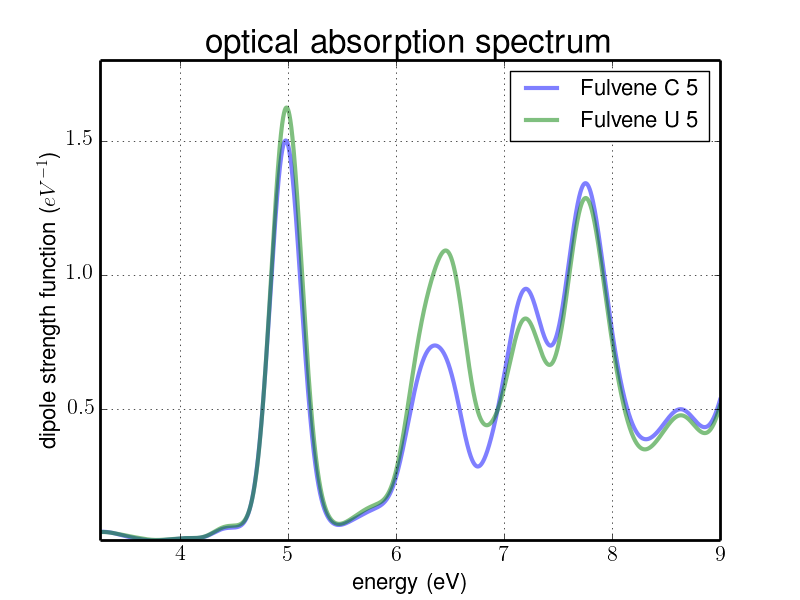}
\caption[]{The electronic spectrum of the fulvene molecule obtained using coupled and uncoupled subsystem rt-TDDFT methods.}
 \label{fig:coupled_uncoupled}
\end{figure}

While reproducing the 
supersystem result is an important feature of any subsystem method, the real interest lies in comparing the optical spectra of the subsystems with the optical spectra of the 
isolated 
entities. As can be seen from Fig.\ \ref{fig:benzene_iso_fde}, where the optical spectra of the isolated fulvene molecule and the fulvene molecule in the dimer are depicted, the 
embedding potential generated by the presence of the benzene molecule results in a shift to lower energies. The shift is present for all excitations, but is especially pronounced 
for the excitation lightly higher than $6$ eV. Furthermore, the embedding potential generated by the presence of the benzene molecule has a clear effect on the intensity of the 
peaks, which is especially pronounced for the excitation at $6.44$ eV in the isolated fulvene spectrum.
An important question is the extent to which the full dynamic response is vital for the reproduction of 
the interaction induced shifts  and changes in the oscillation strength. This question has been broadly discussed for the subsystem formulation of linear response TDDFT. In 
the first formulation of the method, one performs a linear response TDDFT calculation on one subsystem while statically including the density of the other subsystem(s) in the 
embedding potential but restricting the response to the active subsystem only.\cite{casi2004,weso2004}. Several studies have shown that this approach is sufficient in reproducing 
optical spectra\cite{neug2005e,neug2005b}, Raman spectra\cite{neug2005c} induced circular dichroism\cite{neug2005d,neug2006b} and electron-spin-resonance hyperfine 
couplings\cite{neug2005f} even in the presence of hydrogen bonding, as long as there are no explicit couplings in the excitations between the systems.  Later on,  
Neugebauer\cite{neug2007} introduced an approach to include couplings between selected excitations. This effect is clearly seen when comparing the optical spectra of the 
embedded fulvene molecule from 
a coupled and uncoupled calculations, as depicted in Fig.\ \ref{fig:coupled_uncoupled}. In the coupled calculation, both molecules were subjected to a pulse and integrated in 
time, 
with a total density update after each time step. In the uncoupled calculation, the density of the benzene molecule is kept frozen during the full length of the simulation, which 
is equivalent to performing the uncoupled version of the linear response FDE calculation.
One can see that both methods produce very similar results for most excitation energies with the exception of the excitation energy at $6.44$ eV, where there is a difference in 
both the interaction induced shift as the associated oscillator strength. The reason for this difference is that the benzene and fulvene molecules are strongly coupled at this 
frequency\cite{krish2014b}. While the 
uncoupled calculation succeeds in reproducing the results for most excitations from an electronic effect through the presence of the density of the other subsystem in the 
embedding potential, it underestimates the effect generated by the other subsystems for the 
coupled excitations. In this particular case, the interaction induced shift is not as strongly pronounced in the uncoupled calculation as in the coupled calculation and resembles 
more the 
excitation energy in the isolated fulvene molecule. A similar effect is found in the 
dimer with a shorter intermolecular distance of $4$ \AA, where an interaction induced shift was observed in the coupled calculation for the excitation energy at $4.99$ eV in the 
fulvene molecule and $6.78$ eV in the benzene molecule. This interaction induced shift in the uncoupled calculation amounted only to half of 
the value. It is also interesting to note that the coupled calculations produce lower oscillation strengths for the coupled excitations, compared to the uncoupled calculations and 
the isolated molecules. This is compensated by higher oscillation strength values at the uncoupled excitations, in order to satisfy the sum rule of the spectrum.


%
\subsubsection{Transport properties}

An important advantage of rt-TDDFT is the ability to study processes in real time, where the system evolves due to an outside perturbation such as an applied field or an 
excitation, i.e.\ thinks of processes such as electronic conductance, excitation transfer and charge transfer. Since 
subsystem rt-TDDFT has the restriction of constant charge on the subsystems, charge transfer cannot be studied straightforwardly, although it is possible by using the restriction 
to our advantage and calculating electronic couplings between charge-localized, diabatic states\cite{pava2011b}. The real limitation for the study of charge transfer is the lack 
of good quality kinetic energy that can adequately describe overlapping densities
and XC functionals that can reliably describe charge separated states. 
With the exact kinetic energy functional, subsystem DFT would reproduce the total DFT result and 
therefore also a charge transfer reaction. However, this would inevitably result in a delocalization of the subsystem electron densities and while the total electron density would 
%
reproduce the KS-DFT one, unless self-interaction corrected XC functionals are used it would likely be incorrect.

To demonstrate the ability of subsystem rt-TDDFT, we will illustrate the explicit transfer of excitation energy between two coupled systems as discussed by Krishtal et 
al.\cite{krish2014b}. An Na$_4$ cluster, consisting of two Na$_2$ molecules, is a well studied system\cite{hofm2010,huang2014,vasi2002} which couples strongly at the excitation 
energy of $2.18$ eV. For this purpose, two Na$_2$ molecules were placed at a distance of $6.6$ \AA. At the start of the simulation, an electric field in a direction along the 
Na-Na bond is applied to only one of the subsystems, noted as the donor, with a frequency corresponding to an excitation energy of the sodium dimer of $2.18$ eV. Both of the 
subsystems are evolved simultaneous for $100$ fs with a time step of 2 as. The full cluster is placed in a supercell of $22.7 \times 43.5 \times 22.7$ a.u.$^{3}$ to avoid periodic 
interactions. The calculation was performed with the PBE functional\cite{PBEc},  the LC94\cite{LC94} 
functional for NAKE and ultrasoft 
pseudopotentials from the 
GBRV\cite{garri2014} library with a kinetic energy cutoff of $55.0$ Ry and density cutoff of $660.0$ Ry.

\begin{figure}
\centering
 \includegraphics[scale=0.4]{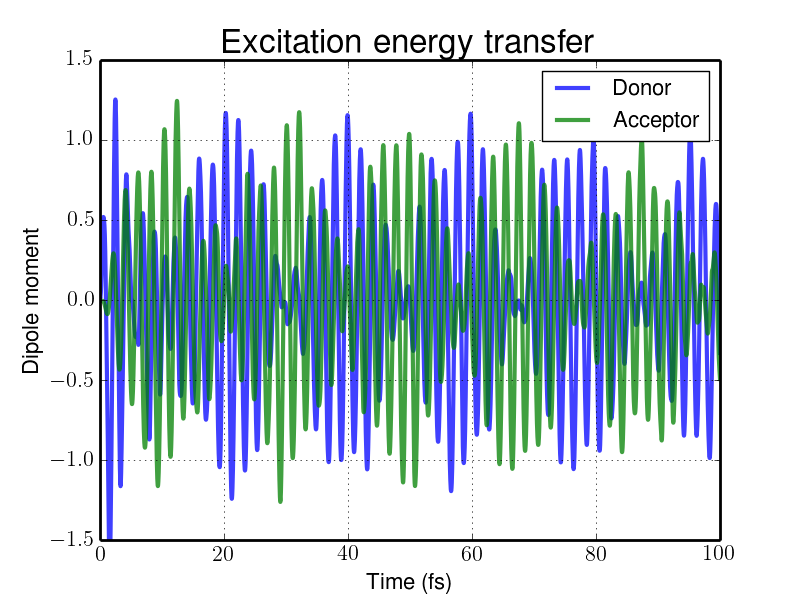}
\caption[]{Excitation energy transfer between two Na$_2$ molecules at the separation distance of $6.6$ \AA, after the excitation of the donor molecule with an applied electric 
field with $2.18$ eV frequency.}
 \label{fig:na4}
\end{figure}

Fig.\ \ref{fig:na4} depicts the evolution of the dipole moment along in 
the direction of the applied field. As one can see, at the beginning of the simulation, the dipole moment of the donor reacts to the applied field and the donor molecule is 
excited. The system proceeds in transferring the excitation energy in a periodic fashion between the donor and the acceptor molecules, as can be clearly seen in Fig.\ 
\ref{fig:na4}. Note that the acceptor subsystem has no direct applied field in its potential and only experiences the excitation due 
the embedding potential. The alternate beating of the dipole moment of the two subsystems as a direct consequence of excitation energy transfer and a full cycle, between the 
maxima of two consequent beatings on the same subsystem, is the rate of the excitation transfer. For the particular separation distance of $6.6$ \AA,  it equals to $19.1$ fs. The 
excitation energy transfer rate is a well studied phenomenon, related to the F\"{o}rster and Dexter energy transfer theories. F\"{o}rster\cite{foer1948,foer1965} described 
excitation energy transfer in a simplified model using a dipole-dipole interaction of the transition densities of the two systems, with an $\frac{1}{R^6}$ dependence. This model is 
valid in the long range where the Coulomb interaction is dominant. For shorter range, Dexter\cite{dext1953} extended the model to include higher multipole order and exchange 
effects. In linear response formalism, excitation energy transfer is studied by calculating the excitation energy transfer couplings explicitly. Needless to say, in order to model 
this effect using subsystem TDDFT both in the linear response and real-time formalisms, a coupled formulation of the method is needed.\cite{neug2010,neug2007}. 

\section{Subsystem prospective of many-body interactions}
\label{manybody}
The calculation of such long-range interactions as van der Waals in molecules pose a major challenge. This is due to the failure of the Local Density Approximation (LDA), 
which makes use of a homogeneous electron gas model, and the Generalized 
Gradient Approximation (GGA), which is semilocal in nature.
Because of this, KS-DFT calculations carried out with semilocal XC functionals is only partially able to account for interactions that are nonlocal in nature, such as dispersion
interactions \cite{mcla1963,long1965} and all long--ranged interactions
originating from the correlated part of the energy functional. However, in those cases where the electron density is nonvanishing in the region separating molecular fragments, semilocal approximations can still effectively account for long-ranged 
interactions due to the correlation energy (see for example Ref.\ \cite{zhao2006}).

In the following, an in-depth analysis of existing \cite{dobs2012} and novel \cite{kevo2014} subsystem formulations of the 
adiabatic-connection fluctuation-dissipation theorem (AC-FDT) in DFT will be described with a particular focus on how these formulations handle density overlap between the subsystems. Practical implementations of these methods as well as related 
ones \cite{Eshuis2011,pode2012,hess2003} will be discussed and preliminary results of the van der Waals inclusive subsystem DFT method will be presented \cite{kevo2014}.

In the AC-FDT, the correlation energy is related to the response functions of the fully-interacting system, $\chi$, and the one of the noninteracting KS system, $\chi_0$. The key is scaling the electron--electron interaction by a coupling constant, 
$\lambda$, in a way that when $\lambda=0$ one recovers the KS system and when $\lambda=1$ the interacting system is recovered \cite{gunn1976,lang1975}. Hence, if one starts from the ground state energy equation for a particular value of $\lambda$, 
given by:
\begin{equation}
E_0^\lambda= \langle \Psi_0^\lambda|\hat{H}^\lambda| \Psi_0^\lambda \rangle
\label{l7}
\end{equation}
it is clear that $E_0^\lambda=E_{KS}$ for $\lambda=0$, i.e., the Kohn-Sham energy and $E_0^\lambda=E_0 $ for $\lambda=1$, i.e., the exact ground state energy of the system.
Thus we get:
\begin{equation}
E_0= E_{KS} + \int_0^1 \d\lambda \frac{d E_0^\lambda}{\d\lambda}.
\label{l8}
\end{equation}
Through the fluctuation-dissipation theorem \cite{call1951}, \eqn{l8} is related to the response functions yielding the following expression for the correlation energy
\begin{equation}
E_{corr}= - \int^1_0 \d \lambda \d\br \d\brp \frac{1}{|\br-\brp|} \int_0^\infty \frac{\d\omega}{2\pi} Im[\chi^\lambda(\br,\brp,\omega) - \chi_0(\br,\brp,\omega)].
\label{ac3}
\end{equation}
According to the prescribed scaling of the electron--electron interaction, $\chi^\lambda$ is given by the Dyson equation:
\begin{equation}
\chi^\lambda(\br,\brp,\omega)= \chi_0(\br,\brp,\omega) + \int \d \bx \int \d \bxp \chi_0(\br,\bx,\omega) \Big{\{} \frac{\lambda}{|\bx-\bxp|} + f^\lambda_{xc}(\bx,\bxp,\omega) \Big{\}} \chi^\lambda(\bxp,\brp,\omega).
\label{l9}
\end{equation}
Thus, van der Waals interactions can be estimated using the AC-FDT in an effective manner \cite{kohn1998}.
\subsection{Adiabatic-connection fluctuation-dissipation theorem for subsystems with nonoverlapping densities}
The Hamiltonian for a model system comprising of two subsystems, $A$ and $B$, could be separated into three components:
\begin{equation}
\hat{H}= \hat{H}_A+ \hat{H}_B+\hat{V}_{AB} 
\label{ac11}
\end{equation}
where $\hat{H}_A$ and $\hat{H}_B$ are the components of isolated subsystems and $\hat{V}_{AB}$ is the interaction which depends on the distance
between them and is given by:
\begin{equation}
\hat{V}_{AB}(\br_1,\br_2) = \frac{1}{|\br_{1}-\br_2|} -\frac{1}{\br_{1B}} -\frac{1}{\br_{2A}}.
\label{ac12}
\end{equation}
Herein, $\br_{1/2}$ belongs to electrons in subsystem 1/2, and $\br_{iX}$ is the electron--nuclear separation between electrons of subsystem $i$ and nuclei $X$.
The Coulomb interaction can be scaled by an adiabatic-connection-like parameter, in the same spirit as AC-FDT. Upon carrying out steps similar to \eqs{l7}{l9}, Dobson \& Gould \cite{dobs2012} derived the nonretarded Lifshitz formula which is 
closely related to the generalized Casimir--Polder equations for two distinguishable
systems (which was derived already with second-order perturbation theory \cite{hess2003,jezi1994}), 
\begin{equation}
E_{\rm disp}= - \frac{1}{2\pi} \int_0^{+ \infty} \d\omega \int \frac{\chi_1(\brp_1,\br_1,i\omega) \chi_2(\br_2,\brp_2,i\omega)}{|\br_1-\br_2||\brp_1-\brp_2|} \d\br_1 \d\br_2 \d\brp_1 \d\brp_2.
\label{ac1}
\end{equation}
Dobson \& Gould made two assumptions: (1) the use of response functions from isolated monomers in the full-potential approximation (i.e.\ the response functions do not depend on the AC-FDT coupling parameter), and (2) that the subsystem's response 
function is given by (in a simplified notation)
\eqtn{ac_dobs}{\chi=\chi_1+\chi_2 + \int \frac{\chi_1\chi_2}{|\br_1-\br_2|}  \d\br_1 \d\br_2.}
The above equation is an approximation because it accounts for the interactions between the response functions of the two subsystems in the RPA approximation, rather than with the full interaction of \eqn{l13}.

\subsection{AC-FDT for density-overlapping subsystems}
In the subsystem DFT scenario, the effective interaction between the subsystem response functions differs from the RPA approximation in the fact that it includes XC and KE terms. This is a pivotal point differentiating the approximate \eqn{ac1} form 
the exact one. For example, the so-called overlap effects are not included in the dispersion energy in \eqn{ac1} \cite{jezi1994}. Neglecting these effects is at the origin of the inclusion of damping factors in common van der Waals corrections to 
semilocal KS-DFT \cite{kris2014}. 

An exact expression for the correlation energy of the interaction between a collection of subsystems can be derived form the subsystem DFT definition of additive and nonadditive
correlation energy functional. Namely,
\begin{equation}
\label{eq:splitcorr}
E_{\rm c}=\underbrace{\sumi E_{\rm c}\left[ \rhoi \right]}_{\mathbf{\mathrm{ Short~Range }}} + \underbrace{E_{\rm c}^{\rm nad}\left[ \rho_I, \rho_{II},\ldots,\rho_{N_S} \right]}_{\mathbf{\mathrm{ Long~Range }}}.
\end{equation}
As the correlation energy is related to the response functions by the AC-FDT, and because of \eqn{sum}, we have that
\begin{equation}
\Delta\chi_\lambda=\chi^\lambda-\chi_0 = \Delta\chi_\lambda^{\rm add} +\Delta\chi_\lambda^{\rm nad}.
\label{f1}
\end{equation}
When the system is separated into smaller subsystems, the coupled response function for each subsystem [$\chi^{c,\lambda}_I$ from \eqn{l17}] and the uncoupled 
response function for each subsystem [$\chi^{u,\lambda}_I$, from \eqn{l18}] must also depend on the value of the coupling constant $\lambda$, and the difference
between these two response functions will, in turn, be related to the correlation effects in each subsystem. The correlation contributions 
from each subsystem should then sum up to give the total correlation of the entire system, which, as has been mentioned already, resides in the
nonadditive component of the $\lambda$ dependent response functions giving rise to the following equation:
\begin{align} 
\nonumber
\Delta\chi_\lambda^{\rm nad} &=\sumi \left(  \chi^{c,\lambda}_I  -
\chi^{u,\lambda}_I \right)\\
\label{django}
&= \sumi \Delta\chi_{\lambda,I}^{\rm nad},
\end{align}
Hence, using the expression for correlation energy, given in Eq.(\ref{ac3}) above, and Eq.(\ref{l17}), we arrive at the following equation:
\begin{equation}
\label{sublr3}
E_{c}^{\rm nad}=-\frac{1}{2\pi}\im\left[\int\int \d \bx_1 \d \bx_2
\frac{ \int_0^1 \d \lambda \int_0^{+\infty} \d \omega 
\sumi \sum_{J\neq I}^{N_S} \chi_I^{u,\lambda}(\omega) K_{IJ}^\lambda(\omega)\chi_J^{c,\lambda}(\omega) }{|\br_1 - \br_2|}\right].
\end{equation}
The above equation achieves the goal of expressing the exact correlation energy of interaction from subsystem quantities (i.e.\ the subsystem response functions) and appropriate kernels of interaction: Coulomb, XC and KE kernels. It generalizes the 
result of Dobson \& Gould for nonoverlapping subsystems in a soft way, i.e.\ \eqn{sublr3} still resembles the generalized Casimir--Polder formula, \eqn{ac1}. 

As all exact equations, it is of impossible evaluation. Thus approximations need to be introduced, and will be discussed in the following section. 

\subsection{Practical implementation}
The nonadditive correlation energy determined with \eqn{sublr3}
is still computationally prohibitive for realistic systems,
as the coupled response functions
must be obtained solving the Dyson-type equation in \eqn{l17}
which in theory requires $N^6$ operations, where $N$ here is the size
of the entire supersystem \cite{pava2013b,koni2012b}. One approximation is achieved
by considering a perturbative solution to the Dyson-type equation \eqn{l17} as
\begin{equation}
\label{sublr4}
\chi_I^{c,\lambda} - \chi_I^{u,\lambda} \simeq \sum_{J\neq I}^{N_S} \chi_I^{u,\lambda} K_{IJ}^{\lambda} \chi_J^{u,\lambda}+\sum_{{ J\neq I,~ K\neq J }}^{N_S} \chi_I^{u,\lambda} K_{IJ}^{\lambda} \chi_J^{u,\lambda} K_{JK}^{\lambda} \chi_K^{u,\lambda}+ 
\cdots
\end{equation}
The above expansion shows how the coupled subsystem response functions contain terms of polarization by other subsystems, which come at the $n$-body level. Retaining only the first term of the expansion (e.g.\ a pair-wise approximation) \eqn{sublr3}
is simplified. The nonadditive correlation energy is now
expressed in terms of $\chi_I^u$ only -- i.e.\ the response polarization is neglected, and only the polarization arising in the ground state is accounted for through the dependence 
of $\chi_I^u$ to the subsystem orbitals.
The next approximation stems from the fact that $K_{IJ}$
is not known exactly. Motivated by the success of RPA for noncovalent
interactions, we approximate $K_{IJ}$ by neglecting the
frequency-dependent exchange-correlation kernel as well as the
kinetic energy contributions to the kernel ($f_{\rm xc}$, $f_{\rm T}$, and $f_{\rm T}^I$)
\begin{equation}
\label{kernel_rpa}
K_{IJ}^{\lambda } \approx 
K_{IJ}^{\lambda, \text{RPA}} = 
 \frac{\lambda}{|\br-\brp|}.
\end{equation}
Since the RPA kernel is frequency independent, it is computationally
much more efficient to solve. Finally,  the last approximation is the full-potential approximation for the response functions.
We summarize the various approximations and their acronyms as proposed in Ref.\cite{kevo2014}
in Table \ref{tappr}.
\begin{table}
\begin{center}
\begin{tabular}{m{3.5cm}cc}
{\sc Approximations} & {\sc Integrand of \eqn{sublr3}} & {\sc Acronym}\\
\hline\\[-8pt]
Exact & $\chi_I^{u,\lambda}K_{IJ}^\lambda\chi_J^{c,\lambda}$ & \\[10pt]
\begin{minipage}{3.5cm}{RPA \\ \& Perturbative}\end{minipage} &
$\chi_I^{u,\lambda}\frac{\lambda}{|\br-\brp|}\chi_J^{u,\lambda}$ &
GCP$_\lambda^u$\\[25pt]
\begin{minipage}{3.5cm}{$\chi_{I/J}^{u,\lambda=1}$\\ \& RPA \\ \&
Perturbative}\end{minipage} &
$\chi_I^{u}\frac{\lambda}{|\br-\brp|}\chi_J^{u}$ & GCP$^{\rm
u}_{1}$\\[30pt]
\begin{minipage}{3.5cm}{nonoverlapping subsystems\\ \& RPA
\\ \& Perturbative}\end{minipage} &
$\chi_I\frac{\lambda}{|\br-\brp|}\chi_J$ & GCP\\[10pt]
\end{tabular}
\end{center}
\caption{\label{tappr} Approximations to \eqn{sublr3} proposed in Ref.\cite{kevo2014} and their respective acronym. ``GCP'' stands for Generalized Casimir-Polder formula.}
\end{table}

For only two interacting subsystems ($N_S=2$),
application of the GCP$^u_{1}$ approximation and subsequent
integration over $\lambda$ (which gives a factor $\frac{1}{2}$) yields
\begin{equation}
\label{subecorr}
E_{\rm c}^{\rm nadd}=-\frac{1}{2\pi}\im\left[\int\int \d \bx_1 \d \bx_2
\frac{\int \d\bx\d\bxp  \int_0^{+\infty} \d \omega \chi_1^{u}(\bx_1,\bx,\omega) \frac{1}{|\br-\brp|}\chi_2^{u}(\bxp,\bx_2,\omega)}{|\br_1 - \br_2|}  \right].
\end{equation}
Clearly, the approximations that led us to \eqn{subecorr} are similar to the ones leading to \eqn{ac1}. Thus, the two equations resemble each other with one key difference, $\chi_I^u\neq \chi_I$. With that, we should expect \eqn{subecorr} to yield somewhat superior dispersion energies than \eqn{ac1}, provided that the ground state density and energy are satisfactorily characterized by the FDE calculation.

When implementing \eqn{subecorr}, for closed--shell systems, the spectral representation of the response functions, assuming real solutions of the TD-DFT eigenvectors, is given by \cite{furc2001}
\begin{eqnarray}
\label{tddft_pc}
\nonumber
\chi_I^u(\br,\brp,\omega) = \sum_{(n)_I} \frac{4\omega_n^u}{(\omega_n^u)^2 -\omega^2}\sum_{(ia)_I,(jb)_I} (X^u_n+Y^u_n)_{ia}(X^u_n+Y^u_n)_{jb} \\
\label{sublr5}
\phi_{i}(\br) \phi_{a}(\br) \phi_{j}(\brp)\phi_{b}(\brp),
\end{eqnarray}
where $\phi_{i}(\br)$ are the KS orbitals of subsystem $I$. Subscripts
$i,j,k,l$ denote occupied orbitals and $a,b,c,d$ virtual orbitals.
$(X^u_n+Y^u_n)_{ia}$ is the projection of the sum of  the excitation
($X$) and de-excitation ($Y$) TD-DFT eigenvector for the $n$-th
excited state of subsystem $I$. The associated eigenvalue is given by
$\omega_n^u$.  The Hartree--XC kernel $K_{II}$
used to determine $\chi_I^u$ is given by \eqn{l13}
with $I=J$ \cite{casi2004,neug2005b}.
Using \eqn{tddft_pc} and employing the GCP$_1^u$
approximation (see Table \ref{tappr}) we can express
the difference of the coupled and uncoupled response functions accordingly, and the AC-FDT formula takes the form
\begin{eqnarray}
\label{sublr8}
\nonumber
E_{\rm c}^{\rm nadd}=\sumi  \sum_{J\geq I}^{N_S} \sum_{(n)_I}^{o_Iv_I}\sum_{(m)_J}^{o_Jv_J} \frac{4}{(\omega_n^u +\omega_n^u)} \sum_{(ia)_I}^{o_Iv_I}\sum_{(jb)_I}^{o_Iv_I}\sum_{(kc)_{J}}^{o_Jv_J}\sum_{(ld)_{J}}^{o_Jv_J}  \\
(X^u_n+Y^u_n)_{ia}(X^u_n+Y^u_n)_{jb} (X^u_m+Y^u_m)_{kc}(X^u_m+Y^u_m)_{ld}\\
\nonumber
\langle ia | ld \rangle \langle jb | kc \rangle.
\end{eqnarray}
The above formula scales as $N_I^3\times N_J^3$ for each pair of subsystems considered. 

A pilot implementation of \eqn{sublr8} has been carried out in a local version of the ADF computer package \cite{adf}, and was presented in Ref.\cite{kevo2014}. The application to a subset of the S22 set gave encouraging results, e.g.\ the binding 
energies of molecular dyads were improved significantly over FDE computed with semilocal nonadditive correlation functionals.

\section{Future directions}
The future of density embedding methods holds promise. This is because partitioning the supersystem into subsystems can be achieved by many different approaches. Thus, the method can capitalize on a plethora of avenues of research. 

In Section \ref{sec:pw}, we have advocated the possibility of using FDE for the description of periodic systems. Recently \cite{lube2014}, the fact that FDE does not impose orthogonality between the subsystem orbitals and yields localized densities 
was exploited to calculate dipole moments of periodic systems using the machinery proper of molecular systems. This provides us with an additional motivation to pursue an ongoing project in our lab which regards devising a subsystem-specific first 
Brillouin zone sampling (the so-called $k$-point sampling). 
This shows that FDE is a very flexible method (e.g.\ the subsystems can be treated at different level of theory). However, when ground states are considered, there are no doubts 
that in order to improve the applicability of FDE, new and more accurate 
nonadditive Kinetic Energy functionals need to be formulated -- a strong overlap between subsystems leads to failure of FDE when semilocal NAKE functionals are employed. Efforts in this direction began with the introduction of nonlocal KE 
functionals \cite{wang2000} and are still ongoing by several groups \cite{lari2014,lari2011,xia2012,xia2012b,bern2008,silv2012,savi2013a}. Despite this, a true breakthrough has still to come for the noninteracting KE functionals.

When departing from the concept of density embedding, the Pauli Blockade (PB) method (reviewed in Section \ref{ort}) provides an elegant solution to the strong-overlap failure of approximate NAKEs. This method, especially when coupled with a 
correlated wavefunction treatment of the subsystems \cite{barn2013,good2012,manb2012}, has already found important applications \cite{good2014}. Other types of embedding include the so-called density-matrix embedding methods \cite{pern2009,kniz2012,
kniz2013,rusa2014} having the advantage of providing an exact embedding framework for uncorrelated (Hartree--Fock, and Kohn--Sham) as well as correlated wavefunctions. The future holds the possibility of using PB for calculating excited states (and 
their properties), as the linear-response theory associated to the PB method has not been developed yet.

The subsystem DFT theory of many-body interactions \cite{kevo2014} presented in Section \ref{manybody} has been implemented in the RPA approximation and using only pair-wise terms in the response function perturbative expansion of \eqn{sublr4}. In 
addition, the full-potential approximation was also used (i.e.\ the response functions are assumed independent on the adiabatic-connection parameter). These choices are very restrictive, as the two-body approximation is known to break down in many 
instances \cite{vlil2010,pode2007}, and the RPA approximation is responsible for overestimated correlation energies \cite{furc2008,kevo2014,eshu2012}. To go beyond the RPA approximation, nonlocal XC and KE functionals are needed \cite{kevo2014,
furc2005,olse2013,olse2014,nguy2009} carrying with them an apparent issue of computational complexity due to the inherent double-integration. Thus, more work is needed to fold in nonlocality in a computationally feasible way, such as by using 
appropriate fitting techniques \cite{roma2009}, ad-hoc corrections \cite{ruzs2011}, or approximating the nonlocal kernel with computationally feasible ones \cite{olse2013,olse2014}. Lifting the full-potential approximation to \eqn{sublr3} will also 
be explored and its effect on the calculated nonadditive correlation energies will be assessed.

One natural question is whether the method arising from \eqn{subecorr} coupled with an FDE calculation of the ground state is capable of reproducing potential energy surfaces (PESs) and not just single point calculations \cite{kevo2014}. It is known 
that certain functionals \cite{pern2009b} can fortuitously reproduce the binding energies in the S22 set without actually being able to reproduce the full PESs. Because the currently available NAKE functionals are unable to describe strongly 
overlapping subsystem densities, we expect this method as implemented with GGA NAKE functionals to fail for short intersubsystem separations and for those dimers which are already not well characterized in their ground states. Work in this direction 
is undergoing in our group, and will also consider the possibility of adding a van der Waals correction to KS-DFT computed with \eqn{subecorr}, using the KS-DFT energy as the uncorrected energy as opposed to the FDE energy.

The real-time TD-DFT implementations (from our group for FDE \cite{krish2014b}, as well as from others for other flavors of subsystem DFT \cite{mosq2013,huang2014}) provide a 
unique avenue for exploring excited states of embedded systems and their 
properties with the possibility of going beyond the linear-response regime \cite{taki2007,taki2008}. Supplementing the dynamics of the electrons with the motion of the nuclei 
leads directly to the formulation of a subsystem nonadiabatic Ehrenfest dynamics. The applications of the rt-TD-DFT code go even beyond this. For example 
correlation energies of interaction [e.g.\ form \eqn{sublr3}] can be 
calculated at imaginary frequency as pioneered by Marques \etal\ \cite{marq2007,oliv2011}, rather than by precalculating the linear response function with the aid of a set of virtual orbitals as customarily done in molecular codes \cite{pode2012,
kevo2014,eshu2010,furc2008}. Due to the time-propagation of the subsystem orbitals, a rt-TD-DFT computation of the nonadditive correlation energy in \eqn{sublr3}, can be done departing from the perturbative pair-wise expansion of the response 
functions. The longer the time propagation, the more the subsystems will couple. This provides us with a very intuitive way to coarse grain in the time-domain which is similar in 
spirit to coarse graining in the imaginary frequency domain \cite{
base2015,pode2012,eshu2010}.

In conclusion, the future looks both bright and busy for the research groups involved in density embedding.

\section*{Acknowledgements}
We would like to acknowledge a number of collaborators with whom we have often interacted during the past two years. Quantum-Espresso guru Dr.\ Davide Ceresoli has helped us in 
achieving the Quantum-Espresso implementations in a finite time and has kindly 
shared with us his own implementation of rt-TDDFT. Interactions with Prof.\ Henk Eshuis have been pivotal in understanding the theory of many-body interactions. We also thank Prof.\ Johannes Neugebauer and Prof.\ Christoph Jacob for fruitful 
discussions. Other group members, Pablo Ramos, Jaren Harrel, and Marc Mankarious, are acknowledged for tolerating the authors during the writing of this work. The donors of the 
American Chemical Society Petroleum Research Fund, and start-up funds 
from Rutgers University--Newark are acknowledged for finantial support. We thank Sven Serneels for proofreading the manuscript.
\bibliographystyle{./aip}
\bibliography{./literature}
\end{document}